\documentclass{mn2e}
\usepackage{graphicx}
\usepackage{amsmath}
\usepackage{amssymb}
\usepackage{algorithm}
\usepackage{algorithmic}

\begin{document}

\title[WiggleZ Survey: cosmic topology]{Using the topology of
  large-scale structure in the WiggleZ Dark Energy Survey as a
  cosmological standard ruler}

\author[Blake et al.]{\parbox[t]{\textwidth}{Chris
    Blake$^1$\footnotemark, J.\ Berian James$^{2,3}$ and Gregory
    B.\ Poole$^4$} \\ \\ $^1$ Centre for Astrophysics \&
  Supercomputing, Swinburne University of Technology, P.O.\ Box 218,
  Hawthorn, VIC 3122, Australia \\ $^2$ Astronomy Department,
  University of California, Berkeley, CA 94720-3411, USA \\ $^3$ Dark
  Cosmology Centre, University of Copenhagen, Juliane Maries Vej 30,
  2100 Copenhagen \O, Denmark \\ $^4$ School of Physics, University of
  Melbourne, Parkville, VIC 3010, Australia}

\maketitle

\begin{abstract}
We present new and accurate measurements of the cosmic
distance-redshift relation, spanning $0.2 < z < 1$, using the topology
of large-scale structure as a cosmological standard ruler.  Our
results derive from an analysis of the Minkowski functionals of the
density field traced by the WiggleZ Dark Energy Survey.  The Minkowski
functionals are a set of statistics which completely describe the
topological nature of each isodensity surface within the field, as a
function of the density value.  Given the shape of the underlying
matter power spectrum, measured by fluctuations in the Cosmic
Microwave Background radiation, the expected amplitudes of the
Minkowski functionals are specified as an excursion set of a Gaussian
random field, with minimal non-Gaussian corrections for the smoothing
scales $\ge 10 \, h^{-1}$ Mpc considered in this analysis.  The
measured amplitudes then determine the cosmic distance $D_V(z)$, which
we obtain with $3-7\%$ accuracies in six independent redshift slices,
with the standard ruler originating in the known curvature of the
model power spectrum at the smoothing scale.  We introduce a new
method for correcting the topological statistics for the
sparse-sampling of the density field by the galaxy tracers, and
validate our overall methodology using mock catalogues from N-body
simulations.  Our distance measurements are consistent with standard
models which describe the cosmic expansion history, and with previous
analyses of baryon acoustic oscillations (BAOs) detected by the
WiggleZ Survey, with the topological results yielding a higher
distance precision by a factor of 2.  However, the full redshift-space
power-spectrum shape is required to recover the topological distances,
in contrast to the preferred length scale imprinted by BAOs, which is
determined by simpler physics.
\end{abstract}

\section{Introduction}
\renewcommand{\thefootnote}{\fnsymbol{footnote}}
\setcounter{footnote}{1}
\footnotetext{E-mail: cblake@astro.swin.edu.au}

The large-scale structure of the Universe, mapped by large galaxy
surveys, is one of the principal tools for testing the physical laws
on cosmological scales; in particular the unknown nature of the `dark
energy' which appears to dominate today's Universe.  The pattern of
the galaxy distribution is sensitive to the matter and energy
constituents of the Universe, the cosmic expansion history, and the
gravitational physics which amplifies the initial density seeds into
today's web of structure.  However, it is also affected by processes
for which there currently exists no complete model: the non-linear
gravitational evolution of structure beyond perturbation theory,
redshift-space distortions due to galaxy peculiar velocities, and
galaxy bias, which describes the complex astrophysical manner in which
the observed galaxy distribution traces the underlying mass.  The
major challenge for cosmological analyses of large-scale structure is
to extract robust information about the underlying cosmological
quantities in the presence of the poorly-modelled non-linear or
astrophysical effects.

For example, one of the most important methods for obtaining robust
cosmological information from large-scale structure surveys is to use
the baryon acoustic oscillations (BAOs) encoded in the clustering
pattern as a standard ruler to map out the cosmic expansion history
(Eisenstein, Hu \& Tegmark 1998, Blake \& Glazebrook 2003, Seo \&
Eisenstein 2003).  This technique exploits a preferred length scale
imprinted in the clustering of galaxies, a late-time signature of the
sound waves which propagated in the pre-recombination Universe
(Peebles \& Yu 1970, Sunyaev \& Zeldovitch 1970, Hu \& Sugiyama 1996).
This preferred scale, accurately calibrated by measurements of the
Cosmic Microwave Background (CMB) radiation, may be extracted from the
galaxy survey observations in a manner which is independent of any
other details of the clustering pattern (e.g., Anderson et al.\ 2012).
This is an attractive approach because the general clustering pattern
is subject to the non-linear and astrophysical distortions mentioned
above (e.g.\ Eisenstein, Seo \& White 2007, Smith, Scoccimarro \&
Sheth 2008, Seo et al.\ 2008), which may be harder to model.  However,
seen from another viewpoint, this `model-independent' technique
excludes information which may in principle be used to improve
cosmological constraints.  The BAO standard-ruler method has now been
applied to a number of galaxy surveys to map out the cosmic expansion
history across a wide range of redshifts (e.g.\ Eisenstein et
al.\ 2005, Percival et al.\ 2010, Blake et al.\ 2011c, Beutler et
al.\ 2011, Anderson et al.\ 2012, Busca et al.\ 2013).

The measurement of BAOs is an example of the use of 2-point clustering
statistics, such as the correlation function or power spectrum, which
are almost ubiquitous for testing cosmological models using galaxy
surveys.  However, the 2-point statistics do not describe all of the
information contained in the cosmological density field.  Lacking any
sensitivity to the phases of the underlying density Fourier modes,
they specifically filter out the direct morphological information
which is most striking in any visual examination of the `cosmic web':
its filamentary nature of inter-connected voids, walls and nodes.
Indeed, two completely different spatial patterns could display the
same 2-point correlation function (e.g.\ Martinez et al.\ 1990).  A
more complete description of the information can make use of a
hierarchy of correlation functions, but these are cumbersome to
implement beyond the 3-point function, and modelling their non-linear
evolution presents difficulties.

A less-studied but promising alternative approach for extracting
information from large-scale structure surveys is to quantify the
topological statistics of the cosmological density field.  In this
study we focus on the Minkowski functionals (Mecke, Buchert \& Wagner
1994), a set of statistics supplied by integral geometry for the
complete morphological specification of spatial patterns.  The
Minkowski functionals are computed from a density field, following
smoothing by a Gaussian filter, by considering the topological nature
of the surfaces formed by each isodensity threshold.  In particular,
for each surface, the four Minkowski functionals describe the volume
enclosed, surface area, curvature and `connectivity' (formally defined
by either the Euler characteristic or genus statistic).  We note that
a number of alternative topological approaches exist for quantifying
large-scale structure such as studies of cosmic voids (Lavaux \&
Wandelt 2012), wavelet analysis (Martinez, Paredes \& Saar 1993),
minimal spanning trees (Barrow, Bhavsar \& Sonoda 1985), and
multiscale morphology filters (Aragon-Calvo, van de Weygaert \& Jones
2010).

Topological statistics are worth exploring as a test of cosmological
models because they may be robust against some of the systematic
non-linear processes which are typically difficult to model in the
correlation functions (Melott, Weinberg \& Gott 1988, Matsubara 2003,
Park \& Kim 2010).  In particular, any process which modifies the
density field, preserving the rank-ordering of density from its
initial state, will not affect the topology of isodensity contours
enclosing a given fraction of volume (Gott, Melott \& Dickinson 1986);
nor does the continuous deformation of a structure affect its
topological connectedness.  As such, the Minkowski functionals are
completely unaffected by linear structure growth and local, monotonic,
non-linear galaxy bias.  Moreover they are only very weakly distorted
by non-linear gravitational evolution and redshift-space distortions
(Matsubara 1994, Matsubara \& Yokoyama 1996).  In summary, the
topology of the density field in co-moving space is exactly conserved
over time during linear evolution, and non-linear corrections remain
small for scales $\ge 10 \, h^{-1}$ Mpc.  Indeed, we determine that
the most important systematic modelling issue in our analysis is not
non-linear evolution, but the `sparse-sampling' distortions arising
when the smoothing scale of the Gaussian filter is comparable to the
mean inter-galaxy separation (James 2012).  We also note that, even if
the initial density statistics were significantly non-Gaussian, the
topological statistics would nonetheless be conserved during linear
evolution.

The pattern of matter overdensities today reflects the distribution of
`seeds' from which they were formed.  If this initial distribution
constituted a Gaussian random field as assumed in this study,
predicted by simple models of inflation, and supported by observations
of the CMB, then the Minkowski functionals of the smoothed density
field have simple analytic forms.  In this case the dependence of the
functionals on the isodensity threshold $\nu$ is a known function of
$\nu$, regardless of the power spectrum of the field, with an unknown
overall normalization that only depends on the shape of the underlying
power spectrum at the smoothing scale.  If the shape of this power
spectrum is known, then theory predicts each of the Minkowski
functionals, independently of the normalization of the underlying
power spectrum.

A measurement of the Minkowski functional amplitudes is then sensitive
to the cosmic distance-redshift relation in two ways, which allow a
`standard ruler' technique to be applied (Park \& Kim 2010, Zunckel,
Gott \& Lunnan 2011).  First, the distance-redshift relation
determines the physical length-scales mapped by the survey, and hence
the amplitudes of the Minkowski functionals in dimensional units.
Secondly, the smoothing scale applied when filtering the density field
in order to perform these measurements assumes a fiducial
distance-redshift relation, and selects a scale in the underlying
model power spectrum to which the measurements are sensitive.  For a
power-law power spectrum these two effects are precisely degenerate,
yielding no sensitivity of the Minkowski functional amplitudes to the
distance scale.  However, if the underlying power spectrum possesses a
curvature which is accurately known, for example using models fit to
CMB observations, then this curvature may be used as a standard ruler
to match the smoothing scale which has been applied to the data.  For
a narrow redshift slice $z$ of a galaxy survey, the resulting
observable is the `volume-weighted' distance $D_V(z)$, identical to
the quantity measured by BAO surveys using the angle-averaged
correlation function.

The aim of our study is to measure these topological statistics using
data from the WiggleZ Dark Energy Survey (Drinkwater et al.\ 2010),
which is one of the largest existing large-scale structure surveys,
and provides a uniquely-long redshift baseline ($0.2 < z < 1$) for
testing the cosmological model.  We use the Minkowski functional
amplitudes to measure the distance-redshift relation $D_V(z)$ and
compare the result to analyses using BAOs (Blake et al.\ 2011c),
validating our techniques using mock galaxy catalogues from N-body
simulations.  In comparison to previous analyses which have focused on
measuring just one of the Minkowski functionals, the genus statistic,
from the Sloan Digital Sky Survey (Gott et al.\ 2009, Choi et
al.\ 2010) or from large N-body simulations (Kim et al.\ 2011), we
implement some new methodological developments: (1) we apply a new
method of estimating the galaxy density field correcting for the
survey selection function; (2) we measure and utilize all Minkowski
functionals rather than just the genus statistic; (3) we advocate and
apply a study of the {\it differential} Minkowski functionals rather
than the integral versions, in order to reduce covariance between
measurements at different values of $\nu$; (4) we measure this
covariance between different density thresholds and functionals, and
propagate this information into our cosmological fits; (5) we
prescribe a method for correcting our measurements for
sparse-sampling.

Our paper is structured as follows: in section \ref{secdata} we
describe the various datasets employed in our analysis, including the
WiggleZ galaxy survey and mock catalogues.  In section \ref{secmink}
we present the Minkowski functional measurements and modelling.  In
section \ref{secamp} we extract the Minkowski functional amplitudes
and their covariances, to which we fit cosmological models in section
\ref{seccosmo}.  We present our conclusions in section \ref{secconc}.

\section{Data}
\label{secdata}

\subsection{The WiggleZ Dark Energy Survey}

The WiggleZ Dark Energy Survey (Drinkwater et al.\ 2010) is a
large-scale galaxy redshift survey of bright emission-line galaxies
over the redshift range $z < 1$, which was carried out at the
Anglo-Australian Telescope between August 2006 and January 2011.  In
total, of order $200{,}000$ redshifts of UV-selected galaxies were
obtained, covering of order 1000 deg$^2$ of equatorial sky.  In this
study we analyzed the same final WiggleZ galaxy sample as utilized by
Blake et al.\ (2011c) for the measurements of BAOs in the galaxy
clustering pattern.  After cuts to maximize the contiguity of the
observations, the sample contains $158{,}741$ galaxies divided into
six survey regions -- the 9-hr, 11-hr, 15-hr, 22-hr, 1-hr and 3-hr
regions.  The survey selection function within each region was
determined using the methods described by Blake et al.\ (2010).

\subsection{The Gigaparsec WiggleZ simulation volume}
\label{secgigglez}

We tested our methodology using data from the Gigaparsec WiggleZ
(GiggleZ) N-body simulation (Poole et al.\ in preparation), a $2160^3$
particle dark matter simulation run in a $1 \, h^{-1}$ Gpc box (with
resulting particle mass $7.5 \times 10^9 \, h^{-1} M_\odot$).  The
cosmological parameters used for the simulation initial conditions
were $[\Omega_{\rm m}, \Omega_{\rm b}, n_{\rm s}, h, \sigma_8] =
[0.273, 0.0456, 0.96, 0.705, 0.812]$.  Bound structures were
identified using {\tt Subfind} (Springel et al.\ 2001), which uses a
friends-of-friends (FoF) scheme followed by a sub-structure analysis
to identify bound overdensities within each FoF halo.  We employed
each halo's maximum circular velocity $V_{\rm max}$ as a proxy for
mass, and used the centre-of-mass velocities for each halo when
introducing redshift-space distortions.

Using the GiggleZ simulation halo catalogues we created one
independent, complete realization of the set of six survey regions
compromising the WiggleZ dataset.  We constructed these mock
catalogues by first selecting a subset of dark matter haloes spanning
a small range of $V_{\rm max}$ around 125 km/s, chosen to possess a
similar clustering amplitude to the WiggleZ galaxies, and
corresponding to halo masses around $10^{12} \, h^{-1} M_\odot$.  We
then subsampled these haloes using the survey selection function in
each region to match the observed number of galaxies.  The GiggleZ
mock catalogues were used for testing the cosmological fits to the
topological statistics for systematic errors, by checking for any
significant deviation between the best-fitting parameters and the
input cosmology of the simulation.

\subsection{Lognormal density field catalogues}

For each WiggleZ survey region we also constructed an ensemble of 400
lognormal realizations using the method described by Blake et
al.\ (2011b).  Lognormal realizations, which are Poisson-sampled from
a density field built from a fiducial power spectrum model, are
relatively cheap to generate and provide a reasonably accurate
description of 2-point galaxy clustering for the linear and
quasi-linear scales important for this analysis.  Work is in progress
to construct a larger set of N-body simulation mock catalogues for the
WiggleZ survey, although this is a challenging computational problem
because the typical dark matter haloes hosting the star-forming
galaxies mapped by WiggleZ have mass $\sim 10^{12} \, h^{-1} M_\odot$,
which (for example) is about $\sim 20$ times lower in mass than a
Luminous Red Galaxy sample.  The lognormal catalogues were subsampled
using the survey selection function in each region to match the
observed number of galaxies.  They were used for determining the
covariance matrix of the topological statistics at different density
thresholds and the sparse-sampling correction, both described in more
detail below.

\subsection{Construction of the smoothed density fields}

\begin{figure*}
\centering
\includegraphics[width=\linewidth]{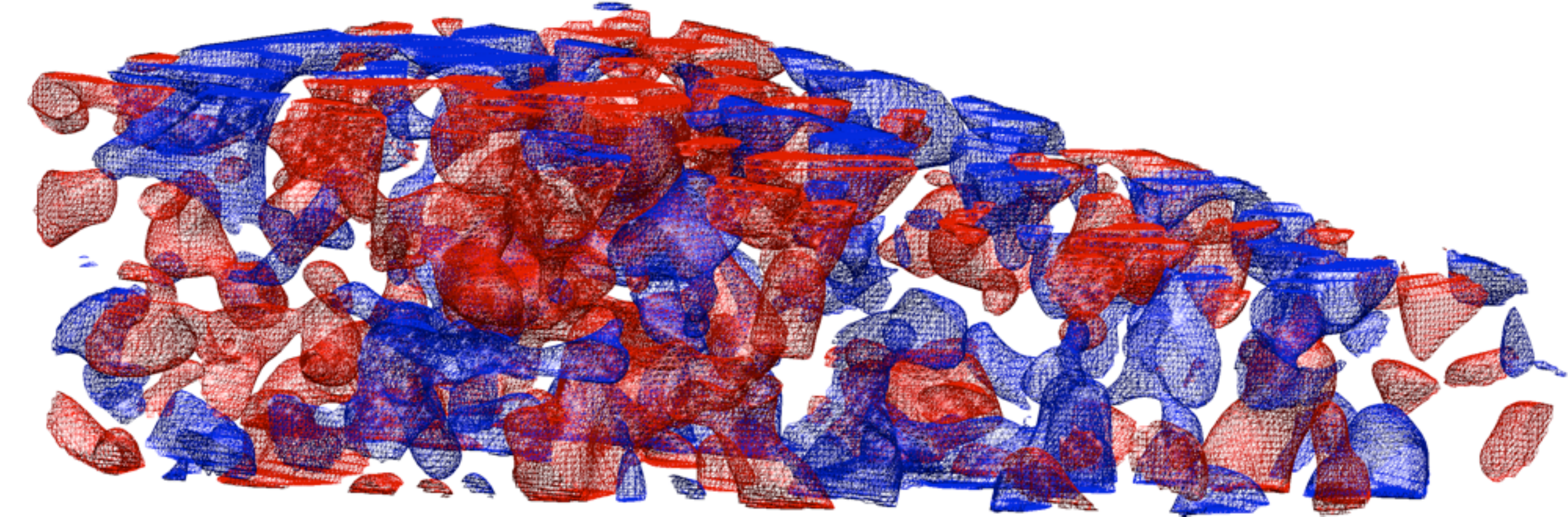}
\caption{Density contours in the WiggleZ survey 15-hr region for a
  smoothing scale of $20 \, h^{-1}$ Mpc, extending from $z=0.3$ (right
  of image) to $z=0.9$, with the contouring chosen to excise the
  highest (red) and lowest (blue) $20\%$ of volume within the field.}
\label{figisodens}
\end{figure*}

The cosmological density field of each dataset was constructed from
the galaxy point distribution by smoothing with a Gaussian filter.  In
the smoothing process we must also correct for the effect of the
varying survey selection function $W(\vec{x})$ with position
$\vec{x}$.  We introduce here a modification to the reconstruction
method used in previous studies (e.g.\ Vogeley et al.\ 1994, James et
al.\ 2009).

The previously existing methodology may be summarized as follows: i)
Each galaxy within the data sample ($D$) is weighted with the value
$1/W(\vec{x})$ and placed in a (padded) three-dimensional array using
a nearest-grid-point binning; ii) these data are smoothed with a
Gaussian ($G$) of standard deviation $R$; iii) the
systematically-lower density near the survey boundaries induced by the
smoothing is characterized by smoothing an array ($C$) that has
constant value inside the survey volume and zero outside; iv) finally,
the smoothed $D$ field is taken in ratio with the smoothed $C$ field,
after which topology of structure within the resulting field can be
studied. Formally, this process may be written:
\begin{equation}
F = [ (D/W) \otimes G ] / [C \otimes G] ,
\end{equation}
where $\otimes$ is used to denote convolution.

The alternative that we propose and implement here is: i) Do not
weight the galaxies initially and instead smooth the galaxy counts in
cells $D$ as they are; ii) instead of creating a comparator field of
constant value inside the survey region, weight the constant field by
the selection function $C \times W(\vec{x})$; iii) smooth this
weighted comparator field, which again is used in ratio with the
smoothed data.  In the notation described immediately above, this
process may be summarized as:
\begin{equation}
F' = [ D \otimes G ] / [ (C \times W) \otimes G ] .
\end{equation}
The motivation for the latter scheme is to apply the selection
function to the data in a smoother and more global way, rather than
locally at the site of each galaxy. In this sense, it is closer in
spirit to the methodology used for correlation function estimation.
These two schemes are identical in the limit that the selection
function does not vary over the scale of the smoothing volume.

Figure \ref{figisodens} shows two isodensity surfaces within the
reconstructed density field of the WiggleZ survey 15-hr region for the
redshift range $0.3 < z < 0.9$, using a Gaussian smoothing scale $R =
20 \, h^{-1}$ Mpc.  The isodensity values have been chosen so as to
excise the highest and lowest density fifths of the field by volume,
and the surfaces display the relative disconnectedness of structure
that is expected for regions this far removed from the median density.
The apparent uniformity of the topology of the structure with redshift
relies on an accurate correction of the effects of the survey
selection function, and the smoothness of the structures themselves is
determined by the choice of filter.

\section{Minkowski functional analysis of galaxy survey data}
\label{secmink}

\subsection{Overview of Minkowski functional methodology}

This work studies the topology of large-scale structure using the four
Minkowski functionals of (the boundary surface of) excursion sets cut
from the density field.  An excursion set is constructed from the
smoothed density field by choosing a critical density threshold
($\rho_c$); regions of density above this value are identified as
being within the surface.  The Minkowski functionals, which we
computed by the algebraic means described in Appendix A, are
identified geometrically with the enclosed volume, surface area,
curvature and genus of the excursion set boundary surface.  Hadwiger's
theorem yields the result that these four statistics form a complete
geometric description of the salient properties of the surface (see
Chen 2004 for a recent review).  Minkowski functionals have been
explored in the context of cosmology by several authors for almost two
decades (early analyses include Mecke et al.\ 1994, Kerscher et
al.\ 1997, Schmalzing \& Buchert 1997).

For convenience we remap the density threshold parameter $\rho_c$ to a
variable $\nu\in(-\infty,\infty)$ which is defined such that the
fraction of volume $V_{\rm frac}$ enclosed by a given isodensity
surface is
\begin{equation}
V_{\rm frac}(\nu) = \frac{1}{\sqrt{2\pi}}\int_\nu^\infty e^{-\nu'^2/2} \,
d\nu' = \frac{1}{2} \, \textrm{erfc} \left( \frac{\nu}{\sqrt{2}} \right)
.
\label{eqnudef}
\end{equation}
This step ensures that the first Minkowski functional -- the enclosed
volume -- is identically an error function irrespective of the
structure of the density field.  Consequently it is the three
remaining functionals, which we studied as a function of $\nu$, that
possess dispositive statistical power in the analysis.

This density transformation is equivalent to the Gaussianisation
process of Weinberg (1992) employed in studies of reconstructing the
linear-regime power spectrum (Neyrinck, Szapudi \& Szalay 2011)
\begin{equation}
\nu \equiv \frac{f_G(\delta)-\bar{f_G}}{\sigma_{f_G}} \;
\textrm{where} \; f_G(\delta) \equiv \textrm{erf}^{-1} \left[
  \int_{-\infty}^\delta f(\delta') \, d\delta' \right] .
\end{equation}
This transformation maps the one-point density distribution
$f(\delta)$ of the field to that of the normal distribution with mean
$\bar{f_G}$ and standard deviation $\sigma_{f_G}$, preserving the
ordering of regions from highest to lowest.  It is also very similar
to the lognormal transformation, given that the cosmological density
field obeys a lognormal distribution even to the smallest scales we
study in this work (Coles \& Jones 1991, Taylor \& Watts 2000, Watts
\& Taylor 2001).

The parameter $\nu$ indexes the surfaces drawn through the density
field.  The first two useful Minkowski functionals describe the area
and curvature of these surfaces.  There is less immediate geometric
intuition for the final functional: the total connectedness, or genus
statistic $g$.  It is defined as the arithmetic difference between the
total number of holes through the filamentary structure and its total
number of disjoint components,
\begin{equation}
g = \textrm{number of holes} - \textrm{number of isolated regions} + 1.
\label{eqgenus}
\end{equation}
The natural interpretation in the context of the cosmological density
field is that the genus number measures how connected (when $g > 0$)
or disjoint ($g < 0$) regions of a given density tend to be. Its
numerical calculation often occurs indirectly via the computation of
the total Gaussian curvature of the surface, a differential geometric
technique introduced by Weinberg, Gott \& Melott (1987).  In this work
we compute an equivalent statistic, the Euler characteristic
[$=4\pi(1-g)$], as the final Minkowski functional.

\subsection{Minkowski functionals of a Gaussian random field}
\label{secgauss}

The cosmological density field may be approximated, when filtered at
certain scales, as a Gaussian random field.  The theory of the
statistics of excursion sets of such fields has been studied by many
authors in contexts of cosmology and geometric statistics
(Doroshkevich 1970; Adler 1981; Bardeen et al.\ 1983, 1986; Hamilton,
Gott \& Weinberg 1986; Tomita 1986; Gott, Weinberg \& Melott 1987;
Ryden 1988; Ryden et al.\ 1989; Matsubara 2003).  For a
three-dimensional Gaussian random field with power spectrum $P(k)$,
smoothed by a Gaussian kernel $G(\vec{x}) =
e^{-(\vec{x}.\vec{x})/2R^2}$ with r.m.s. width $R$, the curves of the
Minkowski functionals $v_n$, in terms of the density parameter $\nu$,
have a known analytical form,
\begin{equation}
v_n[\nu; P(k); R] = A_n[P(k); R] \, e^{-\nu^2/2} \, H_{n-1}(\nu) .
\label{eqgauss}
\end{equation}
Here, values of $n = \lbrace 0,1,2,3 \rbrace$ correspond to the
volume, surface area, mean surface curvature, and Euler
characteristic, respectively, with $H_n$ referring to the Hermite
polynomial of degree $n$
\begin{equation}
H_n(\nu) = e^{\nu^2/2} \, \left( -\frac{d}{d\nu}\right)^n \{
e^{-\nu^2/2} \} ,
\end{equation}
so that 
\begin{equation}
H_0(\nu) = 1; \; H_1(\nu) = \nu; \; H_2(\nu) = \nu^2-1,
\end{equation}
with the extension (Matsubara 2003)
\begin{equation}
H_{-1}(\nu) = \sqrt{\frac{\pi}{2}} \, e^{\nu^2/2} \, \textrm{erfc}
\left( \frac{\nu}{\sqrt{2}} \right) .
\end{equation}
The amplitude functional of the curves, $A_n$, is related to the power
spectrum and smoothing scale in the following manner:
\begin{equation}
A_n\left[ P(k); R \right] = \frac{1}{(2\pi)^{(n+1)/2}} \,
\frac{\omega_3}{\omega_{3-n} \, \omega_n} \, \left(
\frac{\sigma_1^2[P(k); R]}{3 \, \sigma_0^2[P(k); R]} \right)^{n/2} ,
\label{eqamp}
\end{equation}
where $\omega_n = \pi^{n/2}/\Gamma(n/2+1)$ -- in particular, $\omega_0
= 1$, $\omega_1 = 2$, $\omega_2 = \pi$ and $\omega_3 = 4\pi/3$ -- and
the generalized variance functionals are
\begin{equation}
\sigma_j^2[P(k); R] = \frac{1}{2\pi^2} \int k^{2j+2} \, P(k) \,
e^{-k^2 R^2} \, dk .
\label{eqsigsq}
\end{equation}
We note that the dimensions of $A_n$, hence $v_n$, are $({\rm
  length})^{-n}$ -- these equations describe the predicted Minkowski
functionals per unit [length, area, volume] for $n = \lbrace 1,2,3
\rbrace$, matching the dimensions of the estimators described in the
Appendix.

We now provide some intuition for the physical meaning of the ratio of
power spectrum integrals that appears in equation \ref{eqamp}, and for
its dependence on the cosmological distance-scale adopted to analyze
the survey data.  Although in practice we always evaluate the exact
integrals of equation \ref{eqsigsq}, we note that the function
$k^{2j+2} \, e^{-k^2 R^2}$, which weights the power spectrum in the
integrals, peaks at wavenumber $k = \sqrt{1+j}/R$, and hence equation
\ref{eqsigsq} approximately represents the ratio of two power spectrum
amplitudes evaluated at scales $1/R$ and $\sqrt{2}/R$ (i.e.\ between
two wavenumbers in a fixed ratio near $1/R$):
\begin{equation}
A_n \sim C_n \left[ \frac{P(\sqrt{2}/R)}{P(1/R)} \right]^{n/2} ,
\label{eqampapprox}
\end{equation}
where $C_n$ is a constant.  In other words, $A_n$ intuitively depends
on the effective slope of the power spectrum at the smoothing scale.
Assuming a power-law power spectrum $P(k) \propto k^m$ as a concrete
example, we can exactly solve the integrals to find that
\begin{equation}
A_n = C_n \, R^{-n} \, \left( \frac{3+m}{6} \right)^{n/2} .
\label{eqamppow}
\end{equation}

Now suppose we change the distance-scale used to analyze the survey
data, dilating all distances by a factor $\alpha$.  We follow the
normal analysis practice for large-scale structure surveys, keeping
the data measurements fixed and transferring the $\alpha$ dependence
to the model.  For fixed data there are two changes to model: (1) the
smoothing scale $R$ is effectively dilated to $\alpha R$, and (2)
there is an amplitude factor $\alpha^n$ corresponding to the
dependence of the estimators in Appendix A on $({\rm length})^{-n}$.
For the case of a power-law $P(k)$, for which $A_n \propto R^{-n}$, we
can see that these two shifts cancel in equation \ref{eqamppow} such
that the model amplitudes have no dependence on the dilation scale
$\alpha$, hence cannot be used to constrain the distance scale, only
the power-law slope $m$.

However, for a non-power-law $P(k)$, the model Minkowski functional
amplitudes pick up a dependence on the distance scale $\alpha$.  In
the intuitive form used in equation \ref{eqampapprox}:
\begin{equation}
A_n \sim C_n \, \alpha^n \left[ \frac{P(\sqrt{2}/\alpha R)}{P(1/\alpha R)} \right]^{n/2} .
\end{equation}
The `curvature' of the power spectrum $P(k)$ at the smoothing scale
hence provides the `standard ruler' which links the Minkowski
functional amplitudes to the underlying distance scale.

In order to construct the model galaxy power spectrum that appears in
equation \ref{eqsigsq}, we started by generating a matter power
spectrum using the {\tt CAMB} software package (Lewis, Challinor \&
Lasenby 2000).  We assumed the following values for the cosmological
parameters: matter density $\Omega_{\rm m} = 0.27$, Hubble parameter
$h = 0.71$, physical baryon density $\Omega_{\rm b} h^2 = 0.0226$,
primordial spectral index $n_{\rm s} = 0.96$ and normalization
$\sigma_8 = 0.8$, inspired by CMB measurements from the {\it WMAP}
satellite (Komatsu et al.\ 2011); we consider variations of this
fiducial cosmological model in section \ref{seccosmo} below, including
the recent results reported by the {\it Planck} satellite (Planck
collaboration 2013).  We corrected the power spectrum for non-linear
evolution using the `halofit' prescription of Smith et al.\ (2003).
This model power spectrum was subject to further modifications as
described in section \ref{secnongauss}.  We found that using the
`halofit', rather than linear, power spectrum to predict the Minkowski
functional amplitudes was necessary to reproduce the results of the
simulations.

\begin{figure}
\centering
\includegraphics[width=\linewidth]{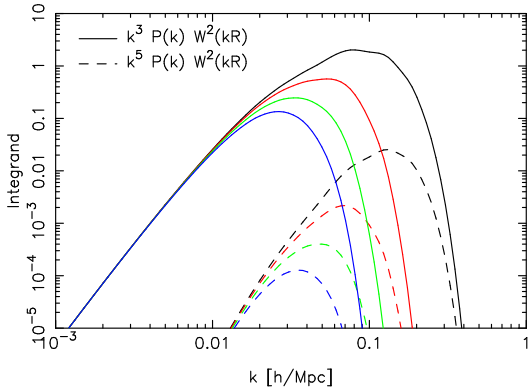}
\caption{The integrands in $k$-space, given our fiducial power
  spectrum model, used in the determination of the amplitudes of the
  Minkowski functionals in equation \ref{eqsigsq}, illustrating the
  range of scales to which these statistics are sensitive.  The solid
  and dashed lines illustrate the integrands of $\sigma_0^2$ and
  $\sigma_1^2$, respectively, and the four sets of curves, from
  top-right to bottom-left, correspond to the four Gaussian smoothing
  lengths $R = (10, 20, 30, 40) \, h^{-1}$ Mpc used in our analysis.
  Noting the logarithmic $y$-axis of the figure, we conclude that our
  measurements are principally sensitive to linear-regime scales $k <
  0.15 \, h$ Mpc$^{-1}$.}
\label{figkwindow}
\end{figure}

Figure \ref{figkwindow} displays the integrand of equation
\ref{eqsigsq} for $j=(0,1)$ as a function of $\ln{k}$, $k^{2j+3} \,
P(k) \, e^{-k^2 R^2}$, for our fiducial power-spectrum model,
illustrating the range of scales to which the topological statistics
are sensitive for the smoothing lengths adopted in our analysis.  We
note that the information is dominated by linear-regime scales $k <
0.15 \, h$ Mpc$^{-1}$.

\begin{figure*}
\centering
\includegraphics[width=14cm]{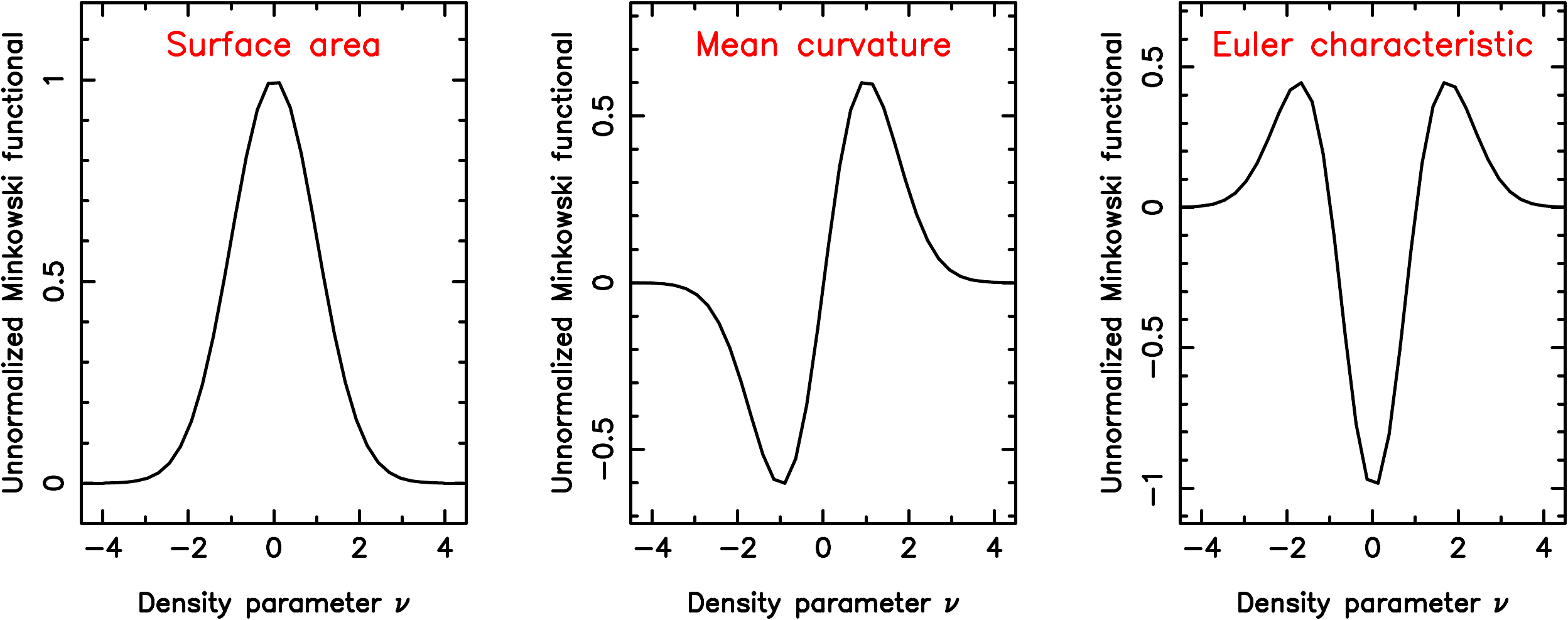}
\caption{The unnormalized shapes of the $n=(1,2,3)$ Minkowski
  functionals for a Gaussian random field, $e^{-\nu^2/2} \,
  H_{n-1}(\nu)$.}
\label{figmfshape}
\end{figure*}

Figure \ref{figmfshape} illustrates the unnormalized shapes of the
$n=(1,2,3)$ Minkowski functionals for a Gaussian random field, with a
view to providing some intuition for the statistics.  We note the
symmetry of these functions for positive and negative $\nu$, such that
the surfaces enclosing overdense and underdense regions possess
similar topological properties.  In the left-hand panel, the area of
these surfaces can be seen to vanish for the highest and lowest
density values as expected, and to peak at average density.  In the
middle panel, the integrated mean curvature of the surfaces also tends
to zero for the highest and lowest density regions, owing both to
their vanishing area and to the fact that at such maxima the surfaces
become spherical, and the sphere is the structure which minimizes
integrated mean curvature.  In the right-hand panel, the Euler
characteristic also approaches zero for the highest and lowest peaks,
given the diminishing number of regions in these limits.  For large
(moderate) departures from the mean density, the surfaces are
preferentially disjoint (connected), corresponding to positive
(negative) Euler characteristic, with a transition at $\nu = \pm 1$
owing to a cancellation between the number of isolated regions and
number of holes specified in equation \ref{eqgenus}.

\subsection{Modifications for non-linear processes}
\label{secnongauss}

\subsubsection{Galaxy biasing}

An attractive property of the Minkowski functionals is that the
density parameter $\nu$ undoes the process of any local, monotonic
galaxy biasing scheme, such that in this model there is no effect of
galaxy bias on the Minkowski functionals (Matsubara 2003), a result
that remains true even in second-order perturbation theory for weakly
non-Gaussian fields (Matsubara \& Yokoyama 1996).  This will not be
the case for non-local or non-deterministic biasing prescriptions,
which we do not consider here.

\subsubsection{Redshift-space distortions}
\label{secrsd}

\begin{figure}
\centering
\includegraphics[width=\linewidth]{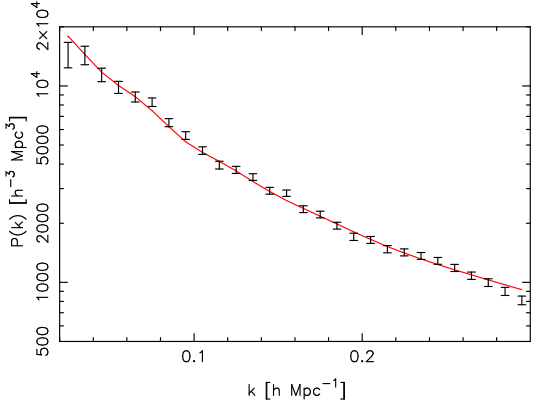}
\caption{The WiggleZ galaxy power spectrum, combining measurements in
  the different survey regions, compared to the model defined in
  sections \ref{secgauss} and \ref{secrsd} which is used to produce
  the amplitudes of the Minkowski functionals.  This model is a good
  description of the data for the range of scales relevant for the
  analysis.}
\label{figpk}
\end{figure}

The observation of galaxies in redshift-space will impart anisotropic
distortions on the power spectrum.  On the large scales relevant to
this analysis, the angle-averaged redshift-space power is given by
\begin{equation}
P(k) = b^2 \, P_{\delta\delta}(k) + \frac{2}{3} b f \,
P_{\delta\theta}(k) + \frac{1}{5} f^2 \, P_{\theta\theta}(k)
\label{eqpkmod}
\end{equation}
(Kaiser 1987) where, in terms of the divergence of the peculiar
velocity field $\theta$, $P_{\delta\delta}(k)$, $P_{\delta\theta}(k)$
and $P_{\theta\theta}(k)$ are the isotropic density-density,
density-$\theta$ and $\theta$-$\theta$ power spectra, and $f$ and $b$
are the cosmic growth rate and galaxy linear bias factor,
respectively.  As discussed above, we produced the matter power
spectrum for our fiducial cosmological parameter set using the
`halofit' model, $P_{\delta\delta} = P_{\rm halofit}$.  We then
generated the velocity power spectra $P_{\delta\theta}$ and
$P_{\theta\theta}$ using the fitting formulae in terms of
$P_{\delta\delta}$, calibrated by N-body simulations, proposed by
Jennings, Baugh \& Pascoli (2011).  We do not include small-scale
velocity dispersion (`fingers-of-god') in our model.  Our
justification of the validity of this model is provided by the tests
we carried out on the N-body simulation mock catalogues, described
below.  We specified fiducial values of $f$ and $b$ as the prediction
of the $\Lambda$CDM growth rate in our fiducial model and the best-fit
to the WiggleZ galaxy 2D power spectra (Blake et al.\ 2011a), noting
that these choices could be varied without significant effect on our
final results.  Figure \ref{figpk} overplots a measurement of the
WiggleZ galaxy power spectrum (Blake et al.\ 2010), combining all
survey regions, and our fiducial power spectrum model, illustrating
that the model provides a good description of the data in the range $k
< 0.3 \, h$ Mpc$^{-1}$ ($\chi^2 = 33.4$ for 27 degrees of freedom).
As quantified further in section \ref{secsparse}, this redshift-space
distortion correction is negligible compared to the statistical errors
in our measurements.

\subsubsection{Non-linear evolution}
\label{secnonlin}

Progress has been made in extending the expressions for the Minkowski
functionals to weakly non-Gaussian fields, particularly those that
depart from gaussianity as a result of primordial physics or
non-linear gravitational evolution.  We summarize here the model we
used for the latter.

The perturbative approach to the study of non-linear gravitational
evolution aims to describe higher-order statistics, such as the
Minkowski functionals, in terms of the lower-order power spectrum.
The Minkowski functionals are expressed in an Edgeworth-like expansion
about the field variance $\sigma_0$, with coefficients derived from
the skewness parameters $S^{(i)}$ (Matsubara 1994, 2003).  To leading
order in $\sigma_0$ this expression reads:
\begin{eqnarray}
 v_n(\nu) &=& A_n \, e^{-\nu^2/2} \, \bigg\lbrace H_{n-1}(\nu) + \sigma_0 \left[ \frac{n}{3} \left( S^{(1)} - S^{(0)} \right) H_n(\nu) \right. \nonumber \\
 && \left.+ \frac{n(n-1)}{6} \left( S^{(2)} - S^{(1)} \right) H_{n-2}(\nu) \right] + {\mathcal O}(\sigma_0^2) \bigg\rbrace .
\label{eqmfmod}
\end{eqnarray}
The skewness parameters are also derived from the power spectrum of
the density field for each smoothing scale $R$:
\begin{eqnarray}
&&
   S^{(0)}(R) =
   (2 + E) S^{11}_0 - 3 S^{02}_1 + (1 - E) S^{11}_2,\\
&&
   S^{(1)}(R) =
   \frac32
   \left[
      \frac{5 + 2E}{3} S^{13}_0 -
      \frac{9 + E}{5} S^{22}_1 -
      S^{04}_1 + \nonumber\right. \\&&\qquad\qquad\qquad\left.
      \frac{2(2-E)}{3} S^{13}_2 -
      \frac{1-E}{5} S^{22}_3
   \right],\\
&&
   S^{(2)}(R) =
   9 
   \left[
      \frac{3 + 2E}{15} S^{33}_0 -
      \frac15 S^{24}_1 -
      \frac{3 + 4E}{21} S^{33}_2 + \nonumber\right. \\&&\qquad\qquad\qquad\left.
      \frac15 S^{24}_3 -
      \frac{2(1-E)}{35} S^{33}_4
   \right],
\end{eqnarray}
where the cosmological factor $E\approx\tfrac{3}{7}$ and, with $l = kR$,
\begin{eqnarray}
&&
   S^{\alpha\beta}_m(R) \equiv
   \frac{\sqrt{2\pi}}{{\sigma_0}^4}
   \left(
      \frac{\sigma_0}{\sigma_1 R}
   \right)^{\alpha+\beta-2}
\nonumber\\
&&\qquad \times\,
   \int \frac{{l_1}^2}{2\pi^2 R^3}
   \frac{{l_2}^2}{2\pi^2 R^3}
   P\left(\frac{l_1}{R}\right) P\left(\frac{l_2}{R}\right)
\nonumber\\
&&\qquad \quad\times\,
   e^{-{l_1}^2-{l_2}^2}
   {l_1}^{\alpha-3/2} {l_2}^{\beta-3/2}
   I_{m+1/2}(l_1l_2) dl_1 dl_2 ,
\end{eqnarray}
where $I_m(x)$ is the modified Bessel function.  Equation
\ref{eqmfmod} specifies the final Minkowski functional model we used
in our analysis, combined with the redshift-space galaxy power
spectrum of equation \ref{eqpkmod}.  As quantified further in section
\ref{secsparse}, this non-linear evolution correction is negligible
compared to the statistical errors in our measurements, partly because
non-linear effects are absorbed by the volume fraction re-mapping of
the density threshold parameter described by equation \ref{eqnudef}
(Matsubara 2003).

\subsection{Differential Minkowski functionals}
\label{secmfdiff}

\begin{figure*}
\centering
\includegraphics[width=16cm]{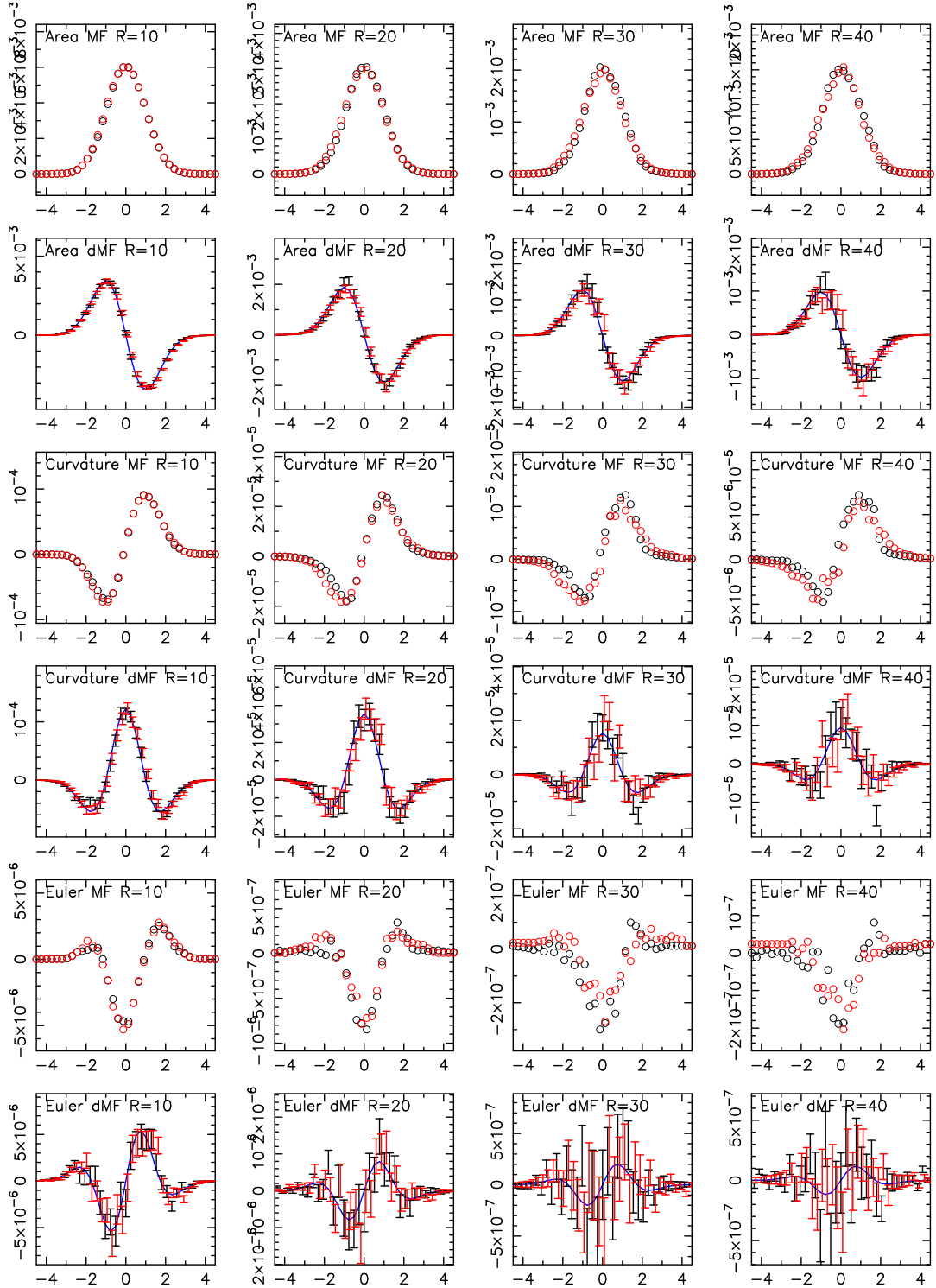}
\caption{Minkowski functional measurements (open circles) and
  differential Minkowski function measurements (error bars) for the $z
  = 0.637$ narrow redshift slice of the WiggleZ 15-hr region for
  smoothing scales $R = 10, 20, 30$ and $40 \, h^{-1}$ Mpc.  The black
  and red symbols represent the WiggleZ data and GiggleZ simulation,
  respectively.  The blue solid line displays the best-fitting model
  in each case, which is a good fit to the data, as discussed in the
  text.  In each panel the $x$-axis represents the density variable
  $\nu$ and the $y$-axis plots the value of the (differential)
  Minkowski functional.}
\label{figmfmeas}
\end{figure*}

The Minkowski functionals, as introduced in the previous subsections
and employed in cosmology to date, possess substantial covariance
between density thresholds that has not been fully detailed in
previous work (although see the appendices of Choi et al.\ (2010) for
recent progress in this endeavour).  Recognising that a substantial
source of this covariance is the use of integral excursion sets, such
that each set is a subset of those that are excised subsequently, we
advance the use of the \emph{differential Minkowski functionals} of
the disjoint part of each subsequent excursion set.  We define the
differential functionals using the algebraic difference between
Minkowski functional measurements at adjacent density thresholds:
\begin{equation}
v_n'(\nu) = \frac{\Delta v_n(\nu)}{\Delta \nu} .
\end{equation}
This is possible because the property of additivity, which the
Minkowski functionals possess, ensures that for the incremental
addition of $\delta S$ to an excursion set $S$:
\begin{eqnarray}
 && v_n(S \cup \delta S) = v_n(S) + v_n(\delta S) - v_n(S \cap \delta
  S) \nonumber \\ & \Rightarrow & v_n(\delta S) = v_n(S \cup \delta S)
  - v_n(S),
\end{eqnarray}
given that $v_n(S \cap \delta S)$ is the null set, since $\delta S$ is
disjoint to the previous surface $S$.  Although the differential
functionals contain no extra information compared to the integral
versions, they result in a more closely-diagonal data covariance
matrix (see section \ref{secmfcov}), which may therefore be estimated
more robustly.

\section{Results}
\label{secamp}

\subsection{Measurement of topological statistics}

We measured the three informative Minkowski functionals (surface area,
curvature, Euler characteristic) of each WiggleZ survey region for
four different Gaussian smoothing scales $R = 10$, $20$, $30$, and $40
\, h^{-1}$ Mpc for 36 values of the density threshold parameter
equally-spaced in the range $-4.5 < \nu < 4.5$, and converted each
measurement to a differential Minkowski functional using a finite
difference.  We split the WiggleZ data into various redshift slices in
the range $0.2 < z < 1$.  First, we performed measurements in broad
overlapping redshift ranges $(0.2 < z < 0.6, 0.4 < z < 0.8, 0.6 < z <
1)$ in order to facilitate comparison with the BAO standard-ruler
distances reported by Blake et al.\ (2011c).  We also split each broad
sample into three narrower, equal-volume redshift slices, such that a
set of independent distance measurements could be constructed in six
narrow redshift slices spanning $0.2 < z < 1$.

We repeated these measurements for the mock halo catalogues
constructed from the N-body simulations.  As described in section
\ref{secgigglez}, we constructed one complete realization of all six
WiggleZ regions for the central broad redshift range $0.4 < z < 0.8$,
which matched the large-scale bias and selection function of the data
sample.  As above, we also split this sample into three narrower,
equal-volume redshift slices spanning this range.

Figure \ref{figmfmeas} displays an example of the integral and
differential Minkowski functional measurements, using the central
narrow redshift slice $(z = 0.637)$ of the range $0.4 < z < 0.8$ of
the 15-hr survey region.  The figure compares the WiggleZ survey
measurement to that determined from the N-body simulation, and the
best-fitting model.  The WiggleZ and simulation results are in good
agreement, and the model is a good fit to the data in all cases, as
judged by the values of the $\chi^2$ statistic.  For the 36 different
combinations of 3 differential Minkowski functionals, 4 smoothing
scales and 3 narrow redshift slices for the range $0.4 < z < 0.8$ of
the 15-hr survey region, the average value of the best-fitting
$\chi^2$ is $34.6$ for the WiggleZ data and $36.5$ for the simulation
data, both for 34 degrees of freedom.

\begin{figure*}
\centering
\includegraphics[width=14cm]{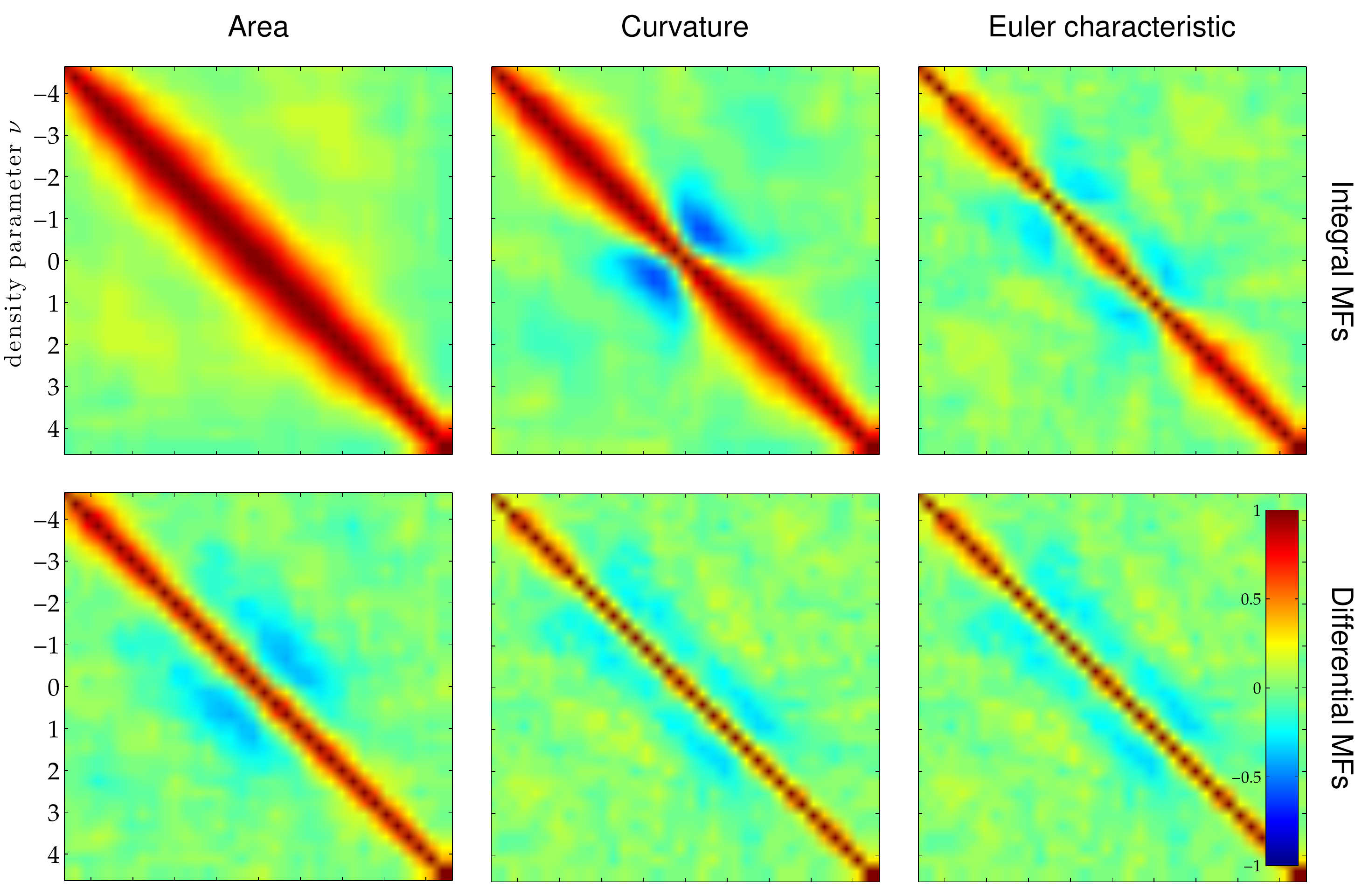}
\caption{Correlation matrices for the three Minkowski functionals and
  their differential forms, generated from 400 realizations of a
  lognormal random field for the 15-hr survey region with the fiducial
  WiggleZ power spectrum, smoothed at a scale of $10 \, h^{-1}$
  Mpc. The colour map has been scaled to the full range $[-1,1]$, so
  that the off-diagonal regions give some indication of the noise
  present in the estimates.}
\label{figmfcov}
\end{figure*}

\subsection{Covariance matrices for differential Minkowski functionals}
\label{secmfcov}

The covariance matrices of the differential Minkowski functional
measurements in each region were determined by measuring these
statistics for each of the ensemble of $N_{\rm log} = 400$ lognormal
realizations.  Writing the measurement at density threshold $\nu_i$ in
the $k$th realization as $v_k(\nu_i)$, the covariance matrices were
determined as
\begin{equation}
C(\nu_i, \nu_j) = \frac{1}{N_{\rm log}-1} \sum_{k=1}^{N_{\rm log}}
\left[ v_k(\nu_i) - \overline{v}(\nu_i) \right] \, \left[ v_k(\nu_j) -
  \overline{v}(\nu_j) \right] ,
\label{eqlog}
\end{equation}
where $\overline{v}(\nu_i) = \sum_{k=1}^{N_{\rm log}} v_k(\nu_i) /
N_{\rm log}$.  The corresponding correlation matrices $C(\nu_i,
\nu_j)/\sqrt{C(\nu_i, \nu_i) \, C(\nu_j, \nu_j)}$ for each statistic
are displayed in figure \ref{figmfcov}, comparing the integral and
differential functionals for the measurements plotted in figure
\ref{figmfmeas}.  This figure explicitly demonstrates that the
covariance matrix of the differential form is more nearly diagonal
than the integral form, as argued in section \ref{secmfdiff}.

\subsection{Correction for sampling systematics}
\label{secsparse}

The measurement of Minkowski functional statistics is systematically
biased in the regime where the smoothing scale is comparable to the
mean inter-galaxy separation (James 2012).  Given the complexity of
the WiggleZ survey selection functions there is no analytic
description of this effect and we relied on an empirical correction
using the lognormal realizations.  We calculated this correction for
each survey region and smoothing scale as the difference between the
mean measured Minkowski functional of the lognormal realizations, and
the Gaussian random-field model Minkowski functional corresponding to
the underlying power spectrum used to generate the lognormal
realizations (which is equivalent to the Minkowski functionals of the
lognormal realization in the limit of high number density).  We note
that this additive correction, which is then applied to each Minkowski
functional measurement from the survey data, is computed independently
of any assumed cosmological model for either the volume of the real
data or the non-linear corrections, but it does assume the fiducial
power spectrum model used to generate the lognormal realizations; it
is beyond the scope of this investigation to consider the
model-dependence of this correction.  The method is verified by its
application to the mock catalogues generated from N-body simulations.

\begin{figure*}
\centering
\includegraphics[width=14cm]{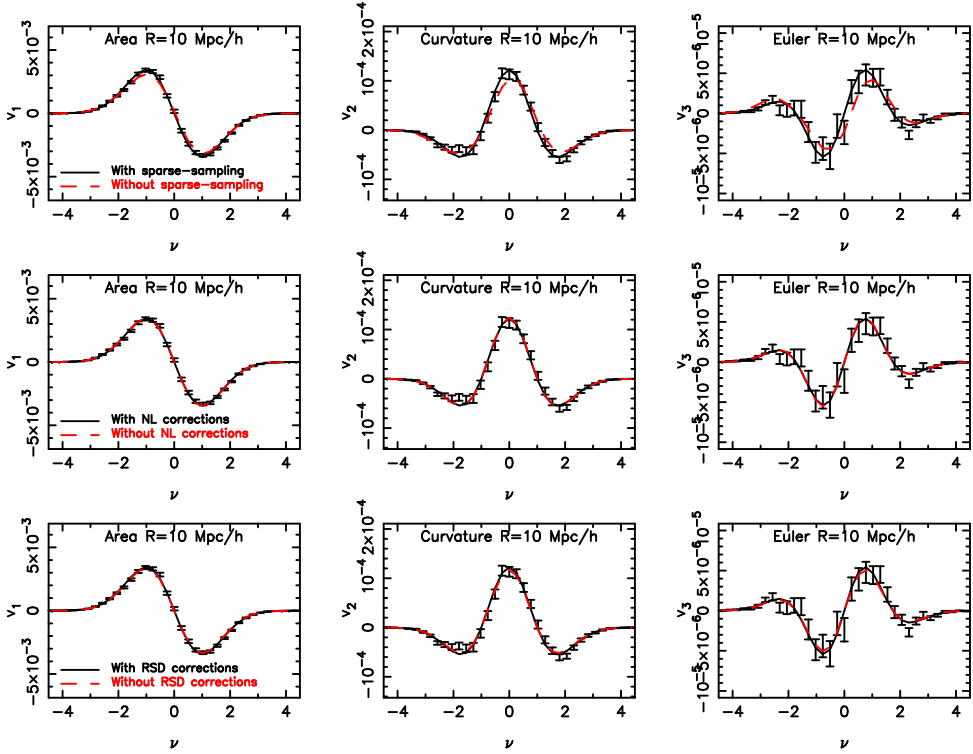}
\caption{The magnitude of the non-gaussian corrections for sparse
  sampling (upper row), non-linear evolution (middle row) and
  redshift-space distortions (lower row), relative to the measured
  differential Minkowski functionals for the most-affected smoothing
  scale, $R = 10 \, h^{-1}$ Mpc.  The panels display results for the
  $z = 0.637$ narrow redshift slice of the 15-hr survey region, with
  the black solid and red dashed line indicating the model with and
  without the application of the non-gaussian correction.}
\label{figsparse}
\end{figure*}

Figure \ref{figsparse} illustrates the relative magnitude of the
sparse-sampling correction (in the upper row) for the most-affected
smoothing scale, $R = 10 \, h^{-1}$ Mpc, again using the example of
the central narrow redshift slice $(z = 0.637)$ of the range $0.4 < z
< 0.8$ of the 15-hr survey region.  For the $10 \, h^{-1}$ Mpc
smoothing scale the correction is comparable to the statistical error
in the measurements; it is negligible for the other smoothing scales
we considered.  Figure \ref{figsparse} also displays the magnitude of
the corrections implied by the non-linear evolution and RSD models
described in section \ref{secnongauss}; these corrections are
negligible in comparison with the statistical errors.

\subsection{Measurements of Minkowski functional amplitudes}

We fitted cosmological models to the Minkowski functional amplitudes,
rather than the functions themselves, given that the amplitudes
contain the cosmological distance-scale information.  We fitted
amplitudes $A_i$ to the differential Minkowski functionals measured
for each WiggleZ survey redshift slice and smoothing scale, using the
covariance matrix determined from the lognormal realizations.  The
Minkowski functional model shapes, $v(\nu)$, were determined as the
random Gaussian field models of section \ref{secgauss} with
corrections for redshift-space distortions (section \ref{secrsd}) and
non-linear evolution (section \ref{secnonlin}).  Due to the
sparse-sampling correction already applied in section \ref{secsparse},
no modelling of shot noise is required.  In order to test for
systematic errors, we repeated these amplitude fits for measurements
from the N-body simulation catalogues.

\begin{figure*}
\centering
\includegraphics[width=14cm]{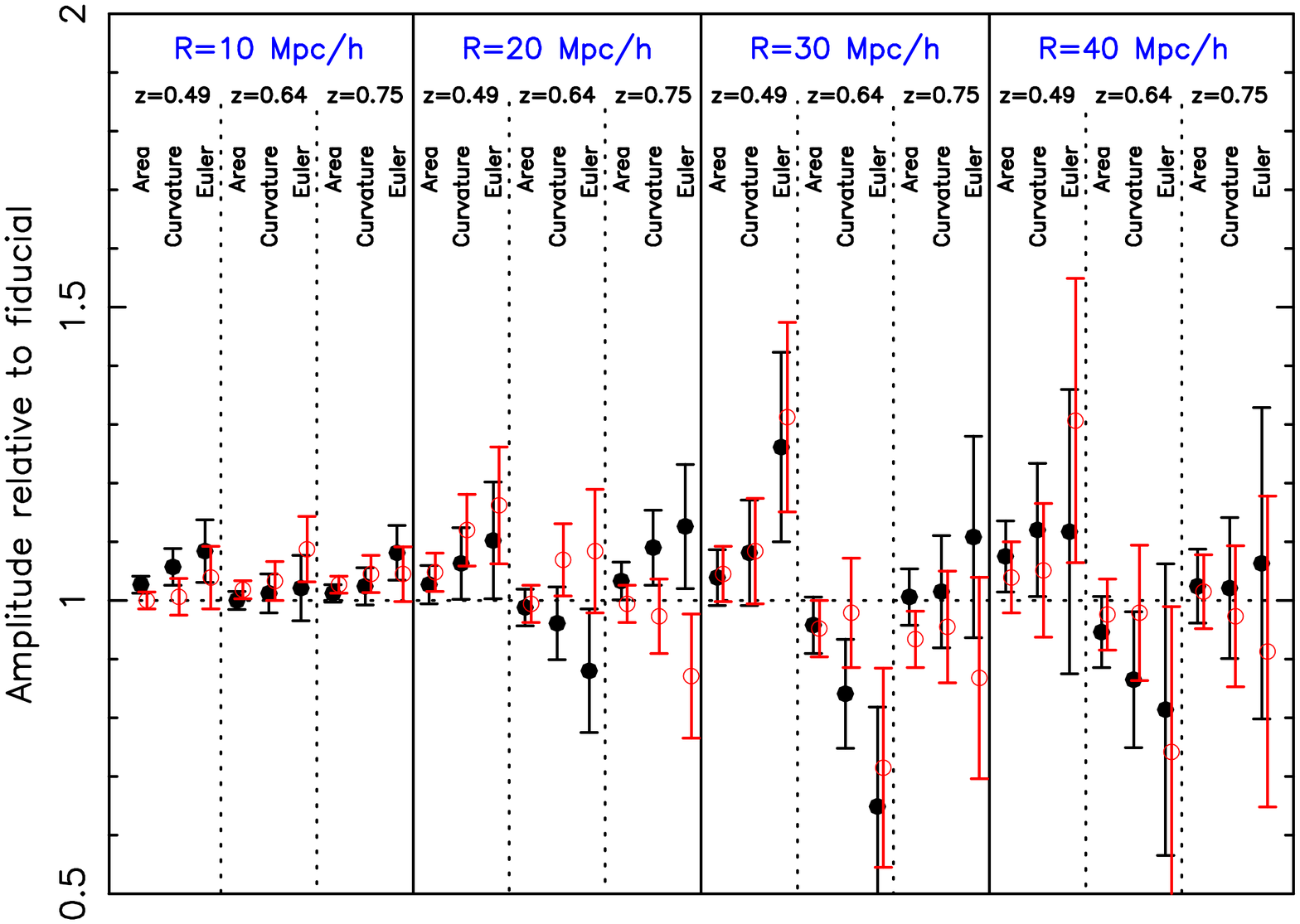}
\caption{Amplitude measurements of each differential Minkowski
  functional for three narrow redshifts slices and four smoothing
  scales $R = (10,20,30,40) \, h^{-1}$ Mpc for the broad redshift
  range $0.4 < z < 0.8$ of the 15-hr survey region.  The solid (black)
  and open (red) circles represent the WiggleZ data and GiggleZ
  simulation, respectively.  The amplitudes are divided by the
  prediction of the fiducial model defined in the text.}
\label{figampmeas}
\end{figure*}

Figure \ref{figampmeas} displays an example of these amplitude fits,
using the broad redshift range $0.4 < z < 0.8$ of the 15-hr WiggleZ
survey region.  There are hence 36 amplitude measurements (spanning 3
Minkowski functionals, 3 narrow redshift slices and 4 smoothing
scales).  For ease of presentation, the results are displayed divided
by the predictions of the fiducial model.  We note the good agreement
between the amplitude measurements of the data and the mock
catalogues, and that the simulation results are consistent with the
GiggleZ input cosmology.

\subsection{Covariance matrix of amplitudes}
\label{secampcov}

\begin{figure*}
\centering
\includegraphics[width=16cm]{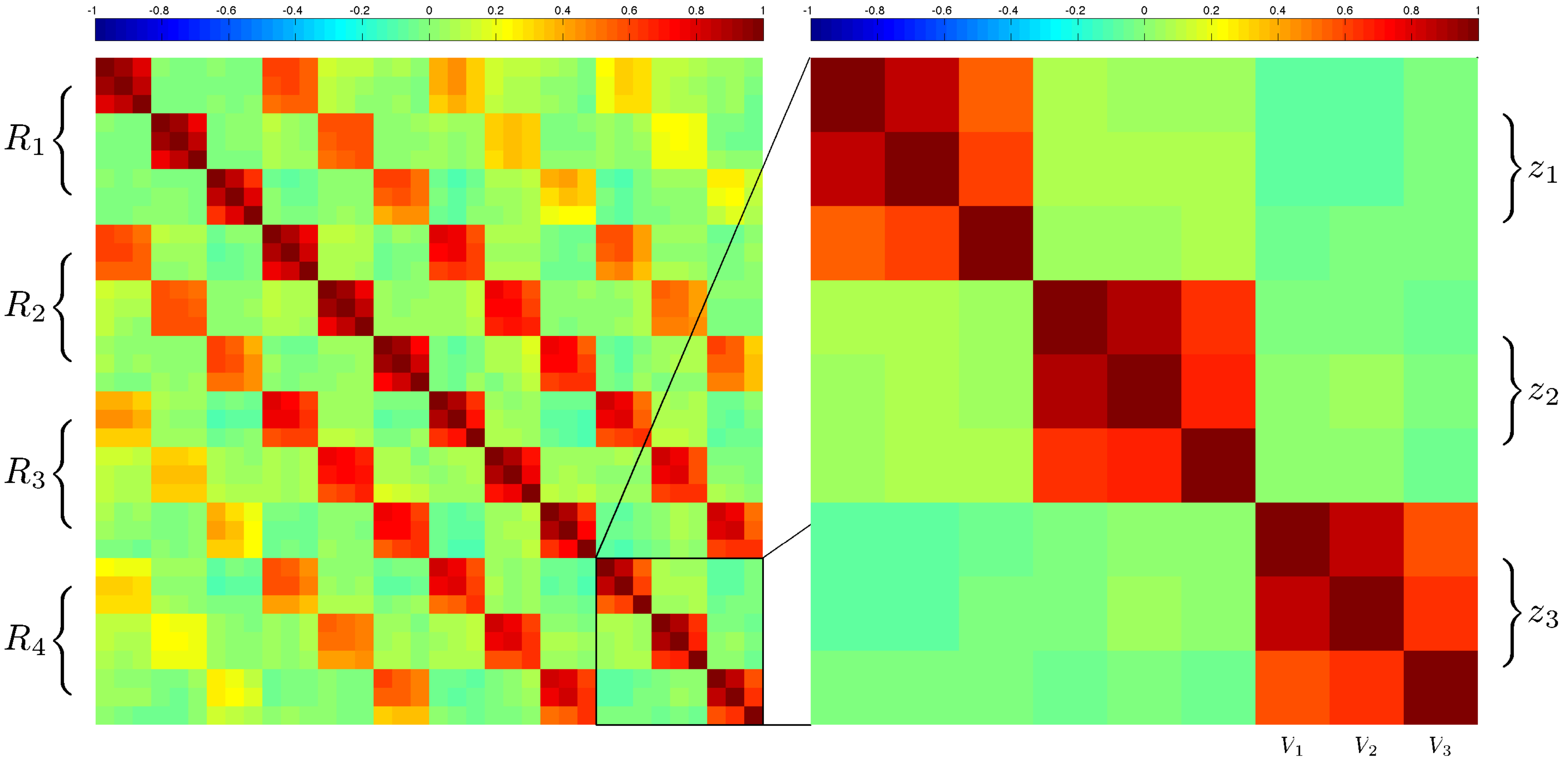}
\caption{The covariance matrix of the amplitude measurements of the 36
  sets of data encompassing combinations of the three Minkowski
  functionals $v_i$, three redshifts $z_i$ and four smoothing scales
  $R_i$ for the broad redshift range $0.4 < z < 0.8$ of the 15-hr
  survey region.  The covariance is displayed as a correlation matrix
  relative to the colour bar at the top of the figure.  The left-hand
  panel displays the full $36 \times 36$ correlation matrix, and the
  right-hand panel is a zoom-in of the lower-right $9 \times 9$
  section corresponding to the largest smoothing length.}
\label{figampcov}
\end{figure*}

The covariance matrix of the amplitude measurements, spanning
different functionals and smoothing scales, was determined by applying
the analysis pipeline described above to every lognormal realization,
and deducing an amplitude covariance matrix using a relation analogous
to equation \ref{eqlog}.  An example amplitude covariance matrix that
results from this process is given in figure \ref{figampcov}, which
displays a $36 \times 36$ matrix corresponding to the measurements in
figure \ref{figampmeas}.  As expected, there are strong correlations
between the amplitudes of different Minkowski functionals measured for
the same redshift interval, and for the same functionals measured for
different smoothing scales.

We note that building the covariance matrix from the lognormal
realizations is a good approximation to the true data covariance.
Considering the diagonal elements, the standard deviation of the
amplitude fits to the lognormal realizations agreed closely with the
standard deviation of the probability distribution obtained when the
amplitudes are fitted to the real survey data.

\begin{figure*}
\centering
\includegraphics[width=13cm]{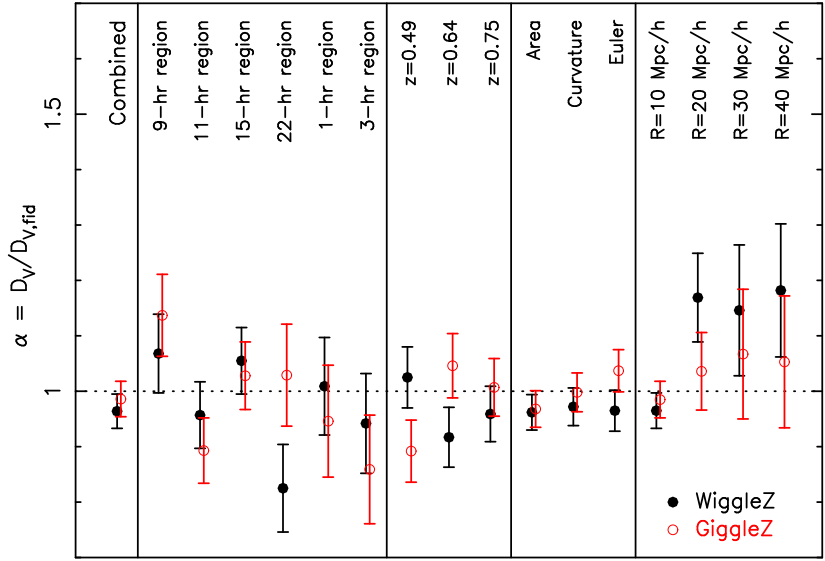}
\caption{The results of distance-scale fits to the set of Minkowski
  functional amplitudes, using the WiggleZ data and simulations for
  the broad redshift range $0.4 < z < 0.8$.  The far-left data point
  displays the measurement that results from combining the set of
  different survey regions, narrow redshift slices, Minkowski
  functionals and smoothing scales.  The subsequent sections of the
  figure, from left to right, restrict the fits to individual regions,
  redshifts, functionals and scales.  The solid (black) and open (red)
  circles show fits to the WiggleZ survey and GiggleZ simulation data,
  respectively.  The fiducial cosmology used to calculate $D_{V,{\rm
      fid}}$ is a flat $\Lambda$CDM model with matter density
  $\Omega_{\rm m} = 0.27$.}
\label{figscalefit}
\end{figure*}

\begin{table*}
\begin{center}
\caption{Distance-scale fits to the topological statisics measured
  from the WiggleZ survey data and GiggleZ N-body simulation mock
  catalogues.  The WiggleZ data is analyzed in a series of broad
  redshift ranges $(0.2 < z < 0.6, 0.4 < z < 0.8, 0.6 < z < 1)$, which
  are also split into 3 narrow, equal-volume redshift subsamples.  The
  mock catalogues were constructed for the range $0.4 < z < 0.8$.  The
  last 4 columns of the table list the effective (volume-weighted)
  redshift $z_{\rm eff}$ of each measurement, the fit of
  $D_V/D_{V,{\rm fid}}$ for fixed power spectrum shape, the value of
  the chi-squared statistic $\chi^2$ for the best-fitting model and
  the number of degrees-of-freedom (`dof'), and the fit of $D_V \,
  \Omega_{\rm m} h^2$ marginalized over $\Omega_{\rm m} h^2$, with
  $D_V$ in units of Mpc.  The fiducial distances, $D_{V,{\rm fid}}$,
  are calculated at each effective redshift assuming a flat
  $\Lambda$CDM cosmological model with matter density $\Omega_{\rm m}
  = 0.27$.  A dataset of six independent distance measurements may be
  constructed using the results corresponding to slices $(1,2,3)$ of
  the $0.2 < z < 0.6$ and $0.6 < z < 1$ redshift ranges.}
\label{tabfit}
\begin{tabular}{ccccccc}
\hline
Sample & Redshift range & slice & $z_{\rm eff}$ & $D_V/D_{V,{\rm fid}}$ & $\chi^2/{\rm dof}$ & $D_V {\rm [Mpc]} \, \Omega_{\rm m} h^2/1000$ \\
\hline
WiggleZ & $0.2 < z < 0.6$ & joint & $0.463$ & $1.02 \pm 0.04$ & $53.1/35$ & $0.213 \pm 0.010$ \\
& & 1 & $0.304$ & $1.16 \pm 0.08$ & $29.4/11$ & $0.191 \pm 0.012$ \\
& & 2 & $0.463$ & $0.96 \pm 0.07$ & $11.3/11$ & $0.205 \pm 0.017$ \\
& & 3 & $0.559$ & $0.97 \pm 0.07$ & $ 6.9/11$ & $0.261 \pm 0.020$ \\
\hline
WiggleZ & $0.4 < z < 0.8$ & joint & $0.637$ & $0.96 \pm 0.03$ & $35.1/35$ & $0.276 \pm 0.009$ \\
& & 1 & $0.486$ & $1.02 \pm 0.05$ & $ 9.7/11$ & $0.250 \pm 0.015$ \\
& & 2 & $0.637$ & $0.92 \pm 0.05$ & $ 9.1/11$ & $0.271 \pm 0.017$ \\
& & 3 & $0.749$ & $0.96 \pm 0.05$ & $13.6/11$ & $0.319 \pm 0.017$ \\
\hline
WiggleZ & $0.6 < z < 1$ & joint & $0.824$ & $0.99 \pm 0.02$ & $41.7/35$ & $0.328 \pm 0.008$ \\
& & 1 & $0.680$ & $1.00 \pm 0.05$ & $ 9.3/11$ & $0.315 \pm 0.016$ \\
& & 2 & $0.824$ & $1.01 \pm 0.04$ & $14.7/11$ & $0.362 \pm 0.014$ \\
& & 3 & $0.944$ & $0.98 \pm 0.03$ & $18.3/11$ & $0.391 \pm 0.014$ \\
\hline
GiggleZ & $0.4 < z < 0.8$ & joint & $0.637$ & $0.99 \pm 0.03$ & $42.2/35$ & $0.280 \pm 0.009$ \\
& & 1 & $0.486$ & $0.89 \pm 0.06$ & $ 7.5/11$ & $0.222 \pm 0.014$ \\
& & 2 & $0.637$ & $1.05 \pm 0.06$ & $16.5/11$ & $0.313 \pm 0.018$ \\
& & 3 & $0.749$ & $1.01 \pm 0.05$ & $13.4/11$ & $0.340 \pm 0.019$ \\
\hline
\end{tabular}
\end{center}
\end{table*}

\begin{figure*}
\centering
\includegraphics[width=13cm]{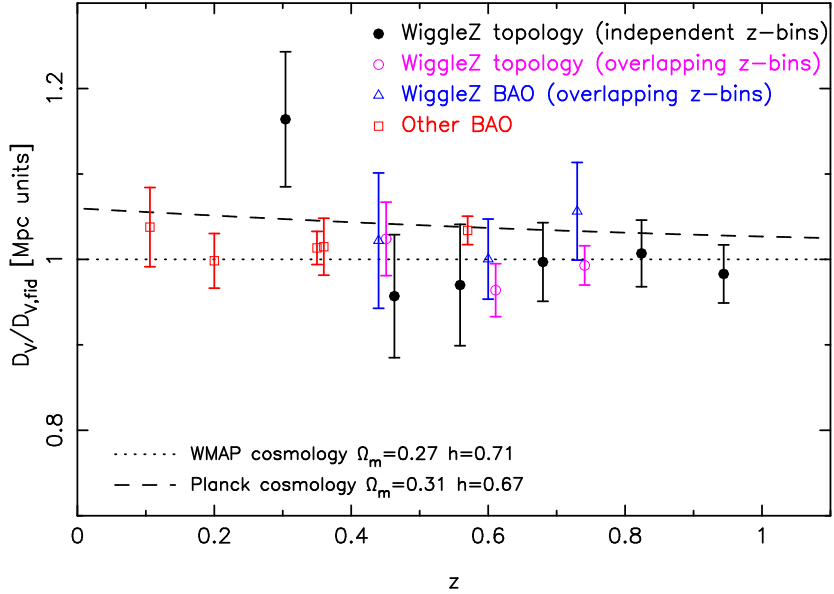}
\caption{Comparison of distance-scale measurements from the
  standard-ruler fits to the WiggleZ topological statistics and recent
  BAO measurements from WiggleZ and other surveys.  The WiggleZ
  topological measurements are shown with two independent binnings:
  the solid black circles are the results for six narrow, independent
  redshift slices spanning $0.2 < z < 1$, and the open black circles
  are the determinations in 3 broader, overlapping redshift ranges.
  These are chosen to coincide with the WiggleZ BAO analysis of Blake
  et al.\ (2011c), whose results are shown as the blue triangles.  BAO
  measurements from other surveys are indicated as red squares, taken
  from Eisenstein et al.\ (2005), Percival et al.\ (2010), Beutler et
  al.\ (2011) and Anderson et al.\ (2012).  The $D_V$ measurements are
  plotted relative to the predictions of the {\it WMAP} fiducial
  cosmological model used in this paper; we also indicate on the
  figure as the dashed line the change relative to this model recently
  implied by the best fits to data from the {\it Planck} satellite
  (Planck collaboration, 2013).  More details about the comparison of
  models and data are given in the text.}
\label{figdvfid}
\end{figure*}

\section{Cosmological model fits}
\label{seccosmo}

\subsection{Distance fits}

Given the shape of the galaxy power spectrum at the relevant smoothing
scales, the cosmological model described in sections \ref{secgauss}
and \ref{secnongauss} prescribes the amplitude of the topological
statistics.  The amplitude measurements were performed using our
fiducial cosmological model, a flat $\Lambda$CDM model with matter
density $\Omega_{\rm m} = 0.27$, to determine the observed survey
distance scale. If the true distance scale deviates from this
fiducial cosmology, which we parameterize by a dilation in distances
by a parameter $\alpha$, then the model amplitudes must be adjusted in
two ways:
\begin{itemize}
\item In the distorted model the dimensional Minkowski functional
  measurements would be scaled by a factor $\alpha^{-n}$, where values
  of $n = \lbrace 1,2,3 \rbrace$ correspond to the surface area, mean
  surface curvature, and Euler characteristic, respectively.  For an
  analysis keeping the data measurement fixed, the model amplitudes
  must therefore be scaled by $\alpha^n$.
\item In the distorted model the true smoothing scales $R$ would have
  changed relative to the fiducial values $R_{\rm fid} = (10, 20, 30,
  40) \, h^{-1}$ Mpc used in the original computation.  The distorted
  scales in the new cosmological model are given by $R = \alpha \,
  R_{\rm fid}$.
\end{itemize}
The dilation scale $\alpha$ is related to the underlying cosmic
distances by $\alpha = D_V(z)/D_{V,{\rm fid}}(z)$, where $D_V(z)$ is
the composite `angle-averaged' distance that is measured using the
baryon acoustic peak in the clustering monopole as a standard ruler
(Eisenstein et al.\ 2005),
\begin{equation}
D_V(z) = \left[ (1+z)^2 D_A(z)^2 \frac{cz}{H(z)} \right]^{1/3} .
\end{equation}
where $D_A(z)$ is the angular diameter distance to the redshift slice
and $H(z)$ is the Hubble expansion parameter.  We can therefore fit
the Minkowski functional amplitude measurements in each redshift slice
for a single value of $\alpha$, hence $D_V(z)$.  We take the effective
redshift $z_{\rm eff}$ of the measurement in each slice as the
volume-weighted redshift of the pixels of the selection function which
are used in the computation.  We fitted the amplitudes after combining
the Minkowski functional measurements in the different survey regions.

An example of these fitting results is displayed in figure
\ref{figscalefit}, combining all survey regions for the $0.4 < z <
0.8$ redshift range, and comparing fits to the WiggleZ survey data and
the GiggleZ N-body simulations.  We show the results both combining
all the different Minkowski functionals and smoothing scales, and
dividing the signal into individual survey regions, narrow redshift
slices, functionals and smoothing scales.  We summarize the
conclusions of figure \ref{figscalefit} as follows:
\begin{itemize}
\item The distance-scale fits to the amplitudes measured from the
  N-body simulation mock catalogues produce results which are
  consistent with the input cosmology of the simulation, validating
  the method.  The measurement of the distance-scale relative to the
  input cosmology of the simulation is $D_V/D_{V,{\rm fid}} = 0.99 \pm
  0.03$ and the value of the $\chi^2$ statistic of the best-fitting
  model is $42.2$ for 35 degrees of freedom.
\item Each Minkowski functional carries roughly equal sensitivity to
  the distance scale, with the area, curvature and Euler
  characteristic producing mutually-consistent distance measurements
  in the redshift range $0.4 < z < 0.8$ with accuracies of $(3.1, 3.4,
  3.6)\%$, respectively.  The accuracy of the combined measurement is
  $2.9\%$, demonstrating that although these topological statistics
  are not independent, their combination does produce a slightly
  improved measurement compared to each individual statistic (in
  particular, an improvement of $20\%$ compared to using the genus,
  i.e.\ Euler characteristic, alone).
\item The fits are dominated by measurements at the smallest smoothing
  scale, $10 \, h^{-1}$ Mpc, which alone produces a $3.2\%$
  distance-scale determination.  The precision resulting from larger
  smoothing lengths is lower, in the range $8 - 12\%$ for $20 - 40 \,
  h^{-1}$ Mpc, due to the smaller effective number of independent data
  samples used to determine the topological statistics as the
  smoothing scale is increased.
\end{itemize}

\subsection{Comparison with BAO distance measurements}

Figure \ref{figdvfid} compiles the overall set of distance-scale
measurements from the fits to the WiggleZ topological statistics, and
compares these with previous measurements of $D_V(z)$ using BAOs as a
standard ruler.  The black, solid circles are the measurements from
WiggleZ topology in six narrow, independent redshift slices spanning
$0.2 < z < 1$, with accuracies in the range $3.3 - 7.7\%$.  The black,
open circles in figure \ref{figdvfid} are the topological measurements
in the broader redshift ranges $(0.2 < z < 0.6, 0.4 < z < 0.8, 0.6 < z
< 1)$, which are designed for comparison with the existing BAO
distance measurements from the WiggleZ survey (Blake et al.\ 2011c).
The two independent techniques for determining the distance scale
produce consistent results, with the topological measurements yielding
a higher accuracy by a factor of 2.  However, the topological
measurements also rely on more assumptions, in particular knowledge of
the shape of the underlying redshift-space galaxy power spectrum,
whereas the BAO technique relies more heavily on the single
standard-ruler scale.  The red squares in figure \ref{figdvfid} are a
compilation of other BAO distance-scale measurements from galaxy
surveys in this redshift range (drawn from Beutler et al.\ 2011,
Eisenstein et al.\ 2005, Percival et al.\ 2010 and Anderson et
al.\ 2012).

In order to place the BAO measurements on figure \ref{figdvfid} we
combined the quoted values of $D_V(z)/r_s(z_d)$ with the latest {\it
  Planck} determination of the sound horizon at the baryon drag epoch,
$r_s(z_d) = 147.4 \times 1.0275$ Mpc (Planck collaboration 2013),
where the factor $1.0275$ converts the exact determination of
$r_s(z_d)$ to the approximation of the Eisenstein \& Hu (1998) fitting
formula used by the BAO papers.  We then divided the result (in Mpc)
by the value of $D_V(z)$ in our fiducial cosmological model, for which
$\Omega_{\rm m} = 0.27$ and $h = 0.71$.  We indicate as the dashed
line in figure \ref{figdvfid} the distances in Mpc relative to this
fiducial model of the cosmological model favoured by {\it Planck},
$\Omega_{\rm m} = 0.31$ and $h = 0.69$ (Planck collaboration 2013),
which provides a somewhat better fit to the BAO dataset, particularly
to measurements from the 6-degree Field Galaxy Survey (Beutler et
al.\ 2011) and the Baryon Oscillation Spectroscopic Survey (Anderson
et al.\ 2012).

The overall picture presented by these measurements is a consistent
delineation of the cosmic distance-scale in the range $z < 1$.  We
note in particular that the WiggleZ topology measurements have
extended this determination into the redshift range $0.8 < z < 1$,
which was not accessible applying the BAO technique to the WiggleZ
survey given that the effective shot-noise weighted cosmic volume it
contains was insufficient to produce a significant detection of the
baryon acoustic peak.  The topological measurements, whose results are
collected in table \ref{tabfit}, do not necessitate a minimum observed
volume.

\subsection{Degeneracy with power spectrum shape}

We now consider the significant degeneracy between the distance-scale
measurements and the shape of the underlying galaxy power spectrum.
For a pure CDM power spectrum, the matter transfer function at
recombination can be expressed as a function of $q = k/\Omega_{\rm m}
h^2$ with $k$ in units of Mpc$^{-1}$ (Bardeen et al.\ 1986).  Given
that changing $D_V$ corresponds to a scale distortion of $k \propto
D_{V,{\rm fid}}/D_V$, we recover that the measured statistics should
depend on the combination $D_V \, \Omega_{\rm m} h^2$ in this
approximation, with $D_V$ in units of Mpc.

\begin{figure}
\centering
\includegraphics[width=\linewidth]{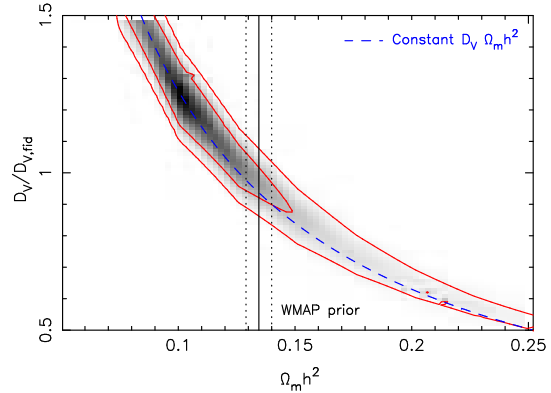}
\caption{The joint probability distribution of $\Omega_{\rm m} h^2$
  and $D_V$ that results from fits to the combined Minkowski
  functionals for the $0.4 < z < 0.8$ redshift range of the WiggleZ
  dataset.  The blue dashed line displays the degeneracy direction of
  constant $D_V \, \Omega_{\rm m} h^2$, which is well-constrained by
  the data.  The vertical solid black line, together with the two
  vertical dotted lines, indicates the best-fit and $\pm 1$-$\sigma$
  range of the measurement from {\it WMAP} (Komatsu et al.\ 2011),
  $\Omega_{\rm m} h^2 = 0.1345 \pm 0.0055$.}
\label{figprob}
\end{figure}

This is illustrated by figure \ref{figprob}, which we generated by
performing a joint fit of $\Omega_{\rm m} h^2$ and $D_V/D_{V,{\rm
    fid}}$ to the WiggleZ topological statistics, where the value of
$\Omega_{\rm m} h^2$ was used to produce the power spectrum model in
each case (with the other cosmological parameters fixed at the values
stated in section \ref{secgauss}) and $D_V/D_{V,{\rm fid}}$ was used
to determine the volume distortion relative to the fiducial cosmology.
As expected, the fits show that there is a significant degeneracy
between these parameters.  The dashed line in figure \ref{figprob}
indicates a set of constant values of $D_V \, \Omega_{\rm m} h^2$,
confirming that this quantity is indeed robustly constrained by the
data, independently of $\Omega_{\rm m} h^2$.  In table \ref{tabfit} we
list the best-fitting values of $D_V \, \Omega_{\rm m} h^2$ for each
data subsample, marginalized over $\Omega_{\rm m} h^2$, which may be
considered more `model-independent' than the measurements of $D_V(z)$,
which assume the fiducial cosmological parameter set.  If we evaluate
the $\chi^2$ values of the `WMAP' and `Planck' $(\Omega_{\rm m}, h)$
models defined above, $(0.27, 0.71)$ and $(0.31, 0.67)$, using the set
of six independent measurements of $D_V \, \Omega_{\rm m} h^2$ in
narrow redshift slices from WiggleZ topology, we find that $\chi^2 =
6.2$ and $13.7$, respectively, for 6 degrees of freedom.  The
corresponding `p-values', indicating the probability of obtaining a
$\chi^2$ equal to these values or higher, are $0.40$ and $0.033$.

We note that the measurements of $D_V$ using topological statistics
are much more precise (by a factor of 3-4) than those which are
obtained by fitting to the shape of the galaxy power spectrum data,
distorting a template model by a scaling factor $\alpha =
D_V/D_{V,{\rm fid}}$ and marginalizing over a normalization factor (as
performed for example in section 4.2 of Blake et al.\ 2011b).  A
possible reason for this is that the Minkowski functionals are
independent of an overall normalization factor such as linear galaxy
bias.

\subsection{Validating the distance error}

\begin{figure*}
\centering
\includegraphics[width=14cm]{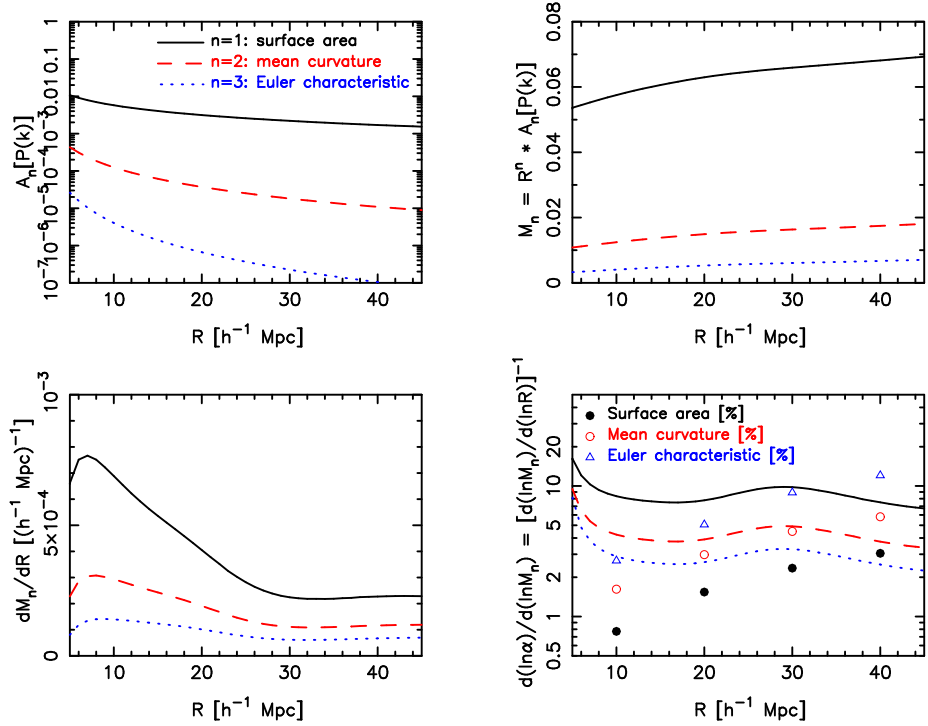}
\caption{An illustration of the step-by-step propagation of the error
  in the Minkowski functional amplitudes $A_n$ to the fitted distance
  scale $\alpha$.  In each panel, the black solid, red dashed and blue
  dotted lines represent functions describing the behaviour of the
  surface area, mean curvature and Euler characteristic, respectively.
  The top-left, top-right, bottom-left and bottom-right panels
  respectively display the dependence on smoothing scale $R = \alpha
  R_{\rm fid}$ of the following quantities: the raw model amplitudes
  $A_n$, the combination $M_n = R^n \, A_n$ which is effectively
  constrained by the data, the rate-of-change $dM_n/dR$ that gives the
  measurement the power to probe the distance scale, and the factor
  $d\ln{\alpha}/d\ln{M_n}$ which maps a fractional error in $M_n$ to a
  fractional error in $\alpha$.  The measured errors in $M_n$ are also
  shown in the bottom-right panel, for the three Minkowski functionals
  and four smoothing scales.  The same $y$-axis range is used, with
  these quantities plotted as a percent error.  More details and
  interpretation are provided in the text.}
\label{figerrtest}
\end{figure*}

Given the impressive accuracy of the distance-scale measurements
provided by these topological statistics, it is important to validate
the plausibility of the errors.  In this section we provide two
supporting arguments.

First, we note that when the distance-scale fits are separately
applied to each of the six individual WiggleZ survey regions that
comprise our dataset, the scatter amongst the best-fitting values is
approximately a factor $\sqrt{6}$ higher than the error in the joint
measurement (as illustrated by the second section of figure
\ref{figscalefit}), providing some approximate empirical verification
of the measurement errors by division of the data into subsets.  We
also split the GiggleZ simulation catalogue into six realizations of
the WiggleZ 15-hr region (as opposed to one realization of all six
WiggleZ regions), and fitted a distance scale to each realization.  We
found that the scatter amongst the best fits was comparable to the
error in the fit to the real 15-hr region dataset (indeed, the scatter
in the simulation results was a little smaller, likely owing to the
fact that the realizations are not truly independent, given that they
have been carved from the same simulation).

Secondly, we demonstrate that the error in the fitted distance scale
can be successfully estimated by propagating the error in the fitted
Minkowski functional amplitudes.  We take the example of the $z =
0.637$ narrow redshift slice, for which we obtain a $5\%$ distance
measurement (see table \ref{tabfit}).  For the surface area, mean
surface curvature and Euler characteristic, the errors in the measured
amplitudes are $(0.8, 1.6, 2.7)\%$ for $R = 10 \, h^{-1}$ Mpc, $(1.5,
3.0, 5.1)\%$ for $R = 20 \, h^{-1}$ Mpc, $(2.4, 4.5, 8.9)\%$ for $R =
30 \, h^{-1}$ Mpc and $(3.1, 5.8, 12.0)\%$ for $R = 40 \, h^{-1}$ Mpc.
As noted in section \ref{secampcov}, these amplitude errors are in
close agreement with the scatter of fits to lognormal realizations.
Furthermore, the Euler characteristic errors agree well with the
$4.2\%$ measurement of the genus amplitude of a similar volume of SDSS
Luminous Red Galaxies for a smoothing scale of $22 \, h^{-1}$ Mpc,
recently presented by Choi et al.\ (2013).

Figure \ref{figerrtest} illustrates the step-by-step propagation of
the error in the Minkowski function amplitudes to the fitted distance
scale $\alpha$.  The top-left panel displays the dependence of the
amplitudes $A_n$ of equation \ref{eqamp} on $R = \alpha R_{\rm fid}$.
This figure gives a falsely optimistic indication of the sensitivity
of the measured amplitudes to $\alpha$; we must also consider the
scaling of the measurements by $\alpha^n$ as the distance scale
changes.  This is encapsulated by the top-right panel, which shows the
variation of $M_n = R^n A_n$ with $R$.  These functions would be
horizontal lines with no dependence on $R$ for a power-law $P(k)$.
The variation of $M_n$ with $\alpha$ controls the propagation of
errors from measured amplitudes to the distance scale, such that the
fractional error in amplitude must be multiplied by a factor
\begin{equation}
\frac{d\ln{\alpha}}{d\ln{M_n}} = \left[ \frac{R}{M_n} \frac{dM_n}{dR}
  \right]^{-1} ,
\end{equation}
to yield the fractional error in the distance scale.  The functions
$dM_n/dR$ are plotted in the lower left-hand panel of figure
\ref{figerrtest}, and the final factors $d\ln{\alpha}/d\ln{M_n}$ are
shown in the lower right-hand panel.  The accuracies of the measured
amplitudes in the $z=0.637$ narrow redshift slice are also displayed
in this panel as a percentage, and it can be seen that multiplying
these accuracies by the relevant factors traced by the lines
successfully reproduces the $\sim 5\%$ distance-scale error.  For
example, error propagation for the measurements with $R = 10 \,
h^{-1}$ Mpc smoothing scale predicts errors in $\alpha$ in the range
$6.4 - 7.7\%$ for the three Minkowski functionals; when combined with
appropriate covariance and added to the (noisier) measurements for
larger smoothing scales, the result is consistent with the final $5\%$
distance error.

\section{Summary}
\label{secconc}

We have presented the first measurements of the cosmic distance scale
using the topology of the galaxy density field as a standard
cosmological ruler.  If the shape of the underlying galaxy power
spectrum is known, then the Minkowski functionals are prescribed via
the statistics of the excursion sets of a Gaussian random field.
Corrections due to non-Gaussian processes are small: the topological
statistics are independent of any local, monotonic, non-linear galaxy
bias and, for the smoothing scales considered in this analysis, are
only weakly distorted by non-linear gravitational evolution and
redshift-space distortions.  As such, the topology of the density
field in co-moving space is exactly conserved during linear evolution
and, given the standard ruler provided by the known curvature of the
underlying power spectrum, may be used to determine the same composite
distance $D_V(z)$ that is probed using baryon acoustic oscillations.

We have applied these techniques to data from the WiggleZ Dark Energy
Survey, implementing a number of methodological improvements compared
to previous analyses:
\begin{itemize}
\item We utilized all Minkowski functionals in our analysis, whereas
  previous work has focused mainly on exploiting the genus statistic.
  Calculating the covariance between the topological statistics, we
  have shown that a combined analysis produces the most accurate
  distance measurements, and that the different statistics provide
  self-consistent results.
\item We studied the differential, rather than integral, Minkowski
  functionals, in order to reduce the covariance between measurements
  at different density thresholds.
\item We employed a series of lognormal realizations, with known
  topological statistics, to determine the correction to the Minkowski
  functionals from the sparse-sampling of the density field by the
  galaxy tracers.  The complexity of the survey selection functions
  implies that this correction does not have an analytic form and must
  be determined numerically.  The ensemble of lognormal realizations
  also provides an accurate covariance matrix of fitted Minkowski
  functional amplitudes, which we used to fit cosmological models.
\end{itemize}
We validated our methodology using mock catalogues sampled from an
N-body simulation, which match both the selection function and
large-scale clustering of the WiggleZ survey data, demonstrating that
the fiducial cosmology of the simulation is recovered (within the
statistical error of the analysis).

When analyzed in broad overlapping redshift ranges $(0.2 < z < 0.6,
0.4 < z < 0.8, 0.6 < z < 1)$, the resulting distance-scale
measurements from the WiggleZ survey have errors in the range $2.1 -
4.1\%$.  These determinations agree with, and are almost twice as
precise as, previous measurements from the same dataset using baryon
acoustic oscillations as a standard ruler.  We used arguments based on
dividing the total dataset into sub-regions, and computing direct
error propagation between the Minkowski functional amplitudes and
distance scale, to increase confidence in the correctness of these
errors.

The topological analysis requires more assumptions, since the full
shape of the underlying power spectrum determines the Gaussian-field
statistics.  We describe this degeneracy by also providing
measurements of the well-constrained combination $D_V \, \Omega_{\rm
  m} h^2$, with errors in the range $2.4 - 4.7\%$.  When analyzed in
six narrow, independent redshift slices in the range $0.2 < z < 1$,
the resulting measurements of $D_V(z)$ have errors in the range $3.3 -
7.7\%$, and agree with the existing set of BAO distance-scale
measurements from other galaxy surveys, and with standard flat
$\Lambda$CDM cosmological models.

We conclude that the utilization of the topological statistics of the
galaxy density field is highly-merited as a complement to standard
analyses based on 2-point statistics, and contains a different set of
systematic errors.  We have demonstrated that these topological
measurements are capable of accurate determinations of the cosmic
distance scale, as advocated by Park \& Kim (2010) and Zunckel et
al.\ (2011).  In the future, topological statistics should also be
useful for distinguishing between different models of gravity (Wang,
Chen \& Park 2012).  Further work is required to model the non-linear
effects of shot noise and redshift-space distortions on these
statistics in a general fashion.

\section*{Acknowledgments}

We are very grateful for an extremely insightful and constructive
referee report from David Weinberg, which greatly improved this paper.
We also appreciated useful feedback on a draft of the paper from
Tamara Davis, Karl Glazebrook, Matthew Colless and Fergus Simpson.
JBJ is grateful for helpful discussions with Taka Matsubara.  CB
acknowledges the support of the Australian Research Council through
the award of a Future Fellowship.  JBJ was supported by the Sophie and
Tycho Brahe Fellowship in Astrophysics jointly between UC Berkeley and
the Dark Cosmology Centre, a position made possible by the Danish
National Research Foundation.  We acknowledge financial support from
the Australian Research Council through Discovery Project grants
DP0772084 and DP1093738 funding the position of GP.  We are also
grateful for support from the Centre for All-sky Astrophysics, an
Australian Research Council Centre of Excellence funded by grant
CE11000102.

{\it Author contributions:} All authors contributed to the development
and writing of this paper.  CB implemented the amplitude and
cosmological fits, and most of the text and figures.  BJ developed the
tools for measuring the density field, the Minkowski functionals,
covariance matrix and amplitudes, and contributed text (including the
Appendix) and figures.  GP generated the N-body simulation catalogues
required for the analysis.

\appendix

\section{\emph{G\'eom\'etrie sans frontieres}, a Minkowski functional measurement implementation for survey data}

We describe an implementation of an algorithm to measure the Minkowski
functionals at density thresholds $v_k(\nu)$ on data with non-periodic
boundaries, as is the case for survey data that has been smoothed and
corrected for selection effects.  Similar enterprises have been
discussed in works two decades past (e.g.\ Coles \& Plionis 1991,
Coles et al.\ 1996), though our approach is novel.  Our routine blends
the {\tt Contour3D} algorithm of Weinberg (1988) with the integral
geometric method for computing the functionals on smoothed fields,
detailed by Schmalzing \& Buchert (1997).  Starting from a
three-dimensional array, it sums the contribution to the four
Minkowski functionals at each array node, where the contribution for
each cell configuration about a node is retrieved from a pre-computed
table. The notable distinction with respect to previous
implementations is that the cell configurations now admit 3 values for
a cell, corresponding to the cell being above the threshold, below the
threshold and not in the survey region. This section describes how the
look-up tables for contributions to the Minkowski functionals are
computed, how the cell configurations are indexed and how the
thresholding and summation is carried out.  We validated our code
using tests on a Gaussian random field generated from the WiggleZ
survey galaxy power spectrum, including selection functions.

The theoretical necessity of an attentive study of boundaries can be
understood from the Gauss-Bonnet theorem. Let $\mathcal{M}$ be a
two-dimensional surface separating regions above and below the
threshold, and $\partial\mathcal{M}$ be the interface between this
surface and regions outside the survey. Then the Euler characteristic
$\chi$, the fourth Minkowski functional, can be evaluated in relation
to the Gaussian curvature $K$ of the surface and the geodetic
curvature $k_g$ along the boundary:
\begin{equation}
\underbrace{\int_{\mathcal{M}} K \, dA}_{=4\pi(1-g)} +
\int_{\partial \mathcal{M}} k_g \, ds = 2 \pi \, \chi(\mathcal{M}).
\end{equation}
The importance of the boundary term is that, when the surface is
closed, the second integral vanishes and the Euler characteristic is
interpreted geometrically as the genus of the surface ($\chi = 2 -
2g$). However, in the case of survey data it is not possible to assert
that the surface is closed---one has no knowledge of the field outside
the survey boundary. The integral geometric algorithm for computing
the Minkowski functionals computes the left-hand side of this
equation, incorporating the boundary term. This produces incorrect
results for the surface area, curvature and Euler characteristic
functionals, a simple demonstration of which is the limit where all of
the survey region is above the density threshold: there are no
interfaces between regions above and below the threshold, so the
Minkowski functionals should have values $\lbrace v_0, v_1, v_2,
v_3\rbrace$ = $\lbrace 1, 0, 0, 0\rbrace$; yet if the computation
includes the boundary term, the latter three functionals will all be
non-zero.

Inevitably, the boundary term must be subtracted, allowing the
functionals to achieve their natural interpretations as volume, area,
curvature and genus and to match the theoretical formulae for these
quantities that have been developed to date.  This is achieved
on-the-fly by requiring that the contributions to the Minkowski
functionals at each node be altered in the presence of boundaries,
which in turn mandates that this implementation operate on a ternary,
rather than binary, threshold array.

\subsection{Computing Minkowski functionals with integral geometry}

To discover how such a computation can be carried out, it is necessary
to return to the fundamentals of Minkowski functional measurement on
smoothed fields in cosmology, explicated most fully in the
dissertation of Jens Schmalzing and summarized in his papers
thereafter (Schmalzing \& Buchert 1997)\footnote{This is an
  appropriate point for us to express our sadness at the tragic and
  untimely passing of Jens Schmalzing in 2005, whose work retains its
  great value to our field through its far-sighted understanding and
  uncompromising clarity of prose.}. Crofton's 1868 formula for
evaluating the length of a curve by counting its intersections with
straight lines drawn through the plane can be extended to the surface
in three dimensions.  This casts the computation of these functionals
as an integral over the intersections between the threshold surface
and all possible hyperplanes. When the threshold surface is embedded
in an array, however, the hyperplanes are those parallel to the
lattice, so that the computation of the geometric properties of the
surface is reduced to combinatorics of the point, line, surface and
cube components of the cells within it.

In the absence of a survey boundary, the threshold surface defines a
binary array of cells above and below the critical
density. Identifying the cells above the threshold as those composing
the volume, the total number of unique vertices $N_0$, edges $N_1$,
faces $N_2$ and cubes $N_3$ within this volume, including those on its
surface, are counted to give the Minkowski functionals (Schmalzing \&
Buchert 1997)
\begin{eqnarray}
v_0 &=& \frac{a^3}{V} N_3  \\
v_1 &=& \frac{a^2}{V}\frac{2}{9}\left(N_2 - 3N_3\right) \\
v_2 &=& \frac{a}{V}\frac{2}{9}\left(N_1-2N_2+3N_3\right) \\
v_3 &=& \frac{1}{V}\left(N_0 - N_1 + N_2 - N_3\right)
\end{eqnarray}
Here, $a$ is the physical length scale of an individual cell and $V$
the physical volume of the field (i.e., $V = a^3N$, where $N$ is the
number of cells in the array), so that the Minkowski functionals $v_k$
are all expressed in physical units and as a fraction of the total
volume of the field. The totals $N_k$ are evaluated by summing the
local contribution $n_k$ at each node across the whole array.

This reduces the computation of the Minkowski functionals to a
counting problem and remains valid even in the presence of survey
boundaries. The crucial change when such a boundary is present is that
the weight assigned to a cell component at the survey interface is
reduced. To discover the manner in which this occurs, consider the
case where the survey volume consists of a single cell of unit size,
above the threshold density, surrounded entirely by unit cells outside
the survey volume. Once again, by geometric argument the Minkowski
functionals of this survey region are $\lbrace v_0, v_1, v_2,
v_3\rbrace$ = $\lbrace 1, 0, 0, 0\rbrace$, yet equations (A1) -- (A4)
will return
\begin{displaymath}
 \lbrace v_{0,1,2,3}\rbrace = \lbrace 1, \tfrac{2}{9}(6 - 3), \tfrac{2}{9}(12-12+3), (8 - 12 + 6 - 3)\rbrace .
 \end{displaymath}
 
This is resolved by reducing the contribution of a cell component to
the count $n_k$ from $1$ to $1 - n_b/2^{3-k}$, where $n_b$ is the
number of cells outside the survey boundary with which the cell
component is in contact. In this particular case, each vertex will now
contribute $1-\tfrac{7}{8}$, each edge $1-\tfrac{3}{4}$ and each face
$1-\tfrac{1}{2}$. When the adjacent cells are not all outside the
survey region, these weightings can take other integer multiples of
$\tfrac{1}{8}$, $\tfrac{1}{4}$ and $\tfrac{1}{2}$ (between 0 and 1)
for vertices, edges and faces respectively. It would not be
uncharitable to characterize as incomplete our understanding of why
this reweighting scheme succeeds. Some further clarifying examples
will be provided in the sections below, after the numerical indexing
scheme is described.

\subsection{Indexing scheme}

\begin{figure}
\centering
\includegraphics[width=\linewidth]{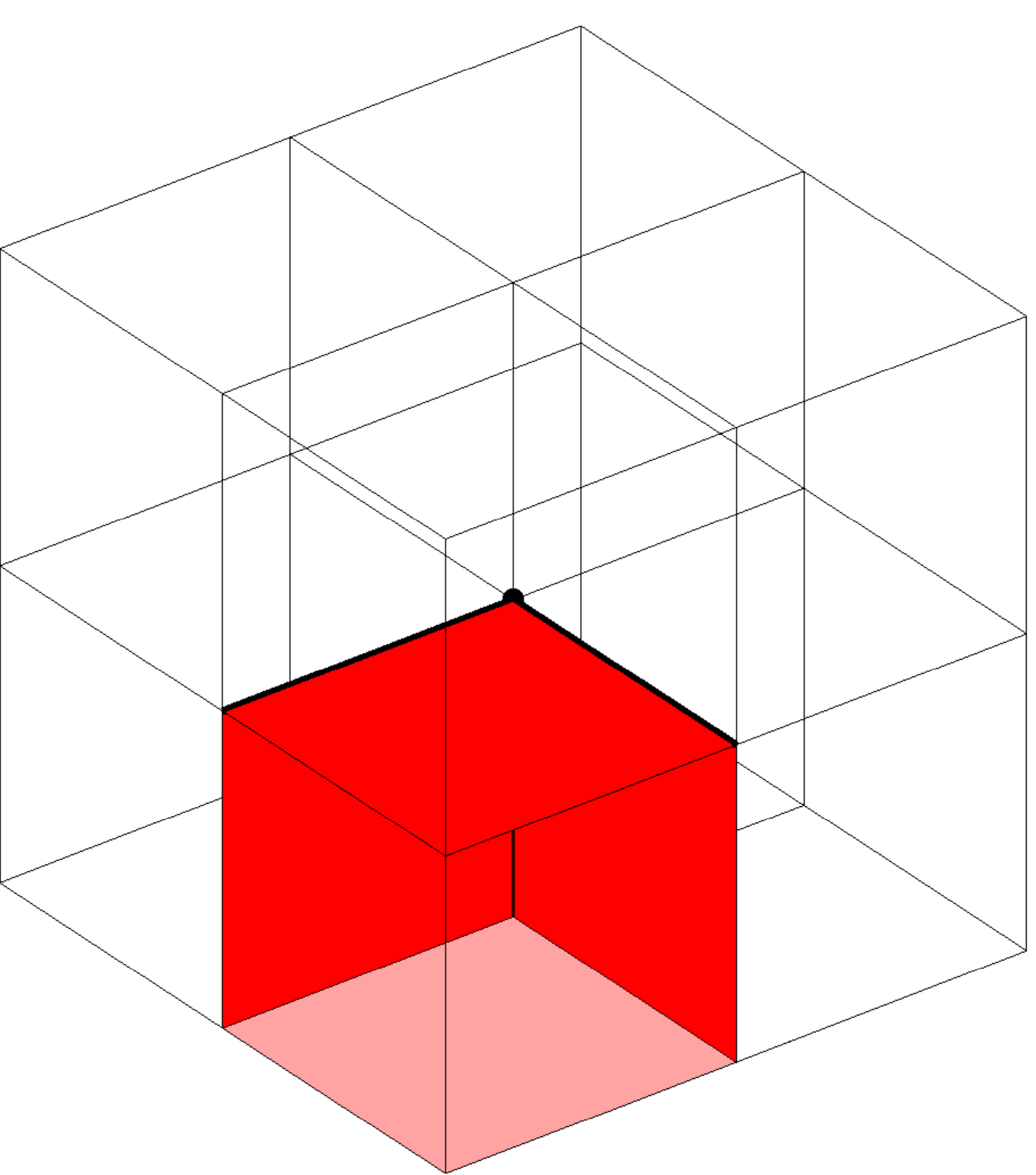}
\caption{Visual description of the node indexing scheme used in this
  algorithm. Cell components are assigned to each array node (centre)
  as follows: one vertex (centre, large point), three edges (thickened
  lines), three faces (opaque shading) and one cube (shaded). This
  scheme ensures that every vertex, edge, face and cube within the
  array will be evaluated exactly once by looping over, or vectorized
  summation of, the array nodes.}
\label{figindexing}
\end{figure}

Computing the number of cell components $n_k$ comprising the
thresholded region, without double counting, can be achieved by
indexing each component uniquely to an array node. An arbitrary node
within the array is surrounded by eight cells, of which one is
assigned to that node; twelve faces, of which three are assigned to
that node; six edges, of which three are assigned to that node; and
one vertex, the node itself. This configuration is demonstrated in
figure \ref{figindexing}. This means that each node can contribute
$\lbrace0,\tfrac{1}{8},\tfrac{2}{8},\ldots,1\rbrace$ vertices,
$\lbrace 0,\tfrac{1}{4},\tfrac{1}{2}\ldots,3\rbrace$ edges,
$\lbrace0,\tfrac{1}{2},1,\ldots,3\rbrace$ faces and
$\lbrace0,1\rbrace$ cubes to each of the totals $n_k$.

\begin{algorithm}
\caption{\texttt{counts} weighted contributions at node}
\label{algnodecount}
\begin{algorithmic}
\REQUIRE a $2^3$ ternary array, $A$
\STATE $n_0 \gets \textbf{or} \left( A[:,:,:] = 1 \right) \times \big( 1-\tfrac{1}{8} \Sigma (A[:,:,:]=2) \big)$
\STATE $n_1 \gets \textbf{or} \left( A[1,:,:] = 1 \right) \times \big( 1-\tfrac{1}{4} \Sigma (A[1,:,:]=2) \big) + \textrm{prms.}$
\STATE $n_2 \gets \textbf{or} \left( A[1,1,:] = 1 \right) \times \big( 1-\tfrac{1}{2} \Sigma (A[1,1,:]=2) \big) + \textrm{prms.}$
\STATE $n_3 \gets \left(A[1,1,1] = 1\right)$
\end{algorithmic}
\end{algorithm}

Given this specification, algorithm \ref{algnodecount} evaluates the
number of vertices, edges, face and cubes, with proper weightings for
survey boundaries, at an individual node. In this notation, `$=$' is
used in the sense of logical evaluation, returning $1$ when the
condition is true, and returning a binary array of the same size as
the object on the left-hand side; \textbf{or} $(X)$ evaluates true when
any member of the array slice $X$ is true; and $\Sigma(X)$ is the sum
over an array slice. Each of the three faces and edges indexed to the
node are tested separately and these permutations have been suppressed
in the expression for $n_1$ and $n_2$.

\begin{figure}
\centering
\includegraphics[width=0.49\linewidth]{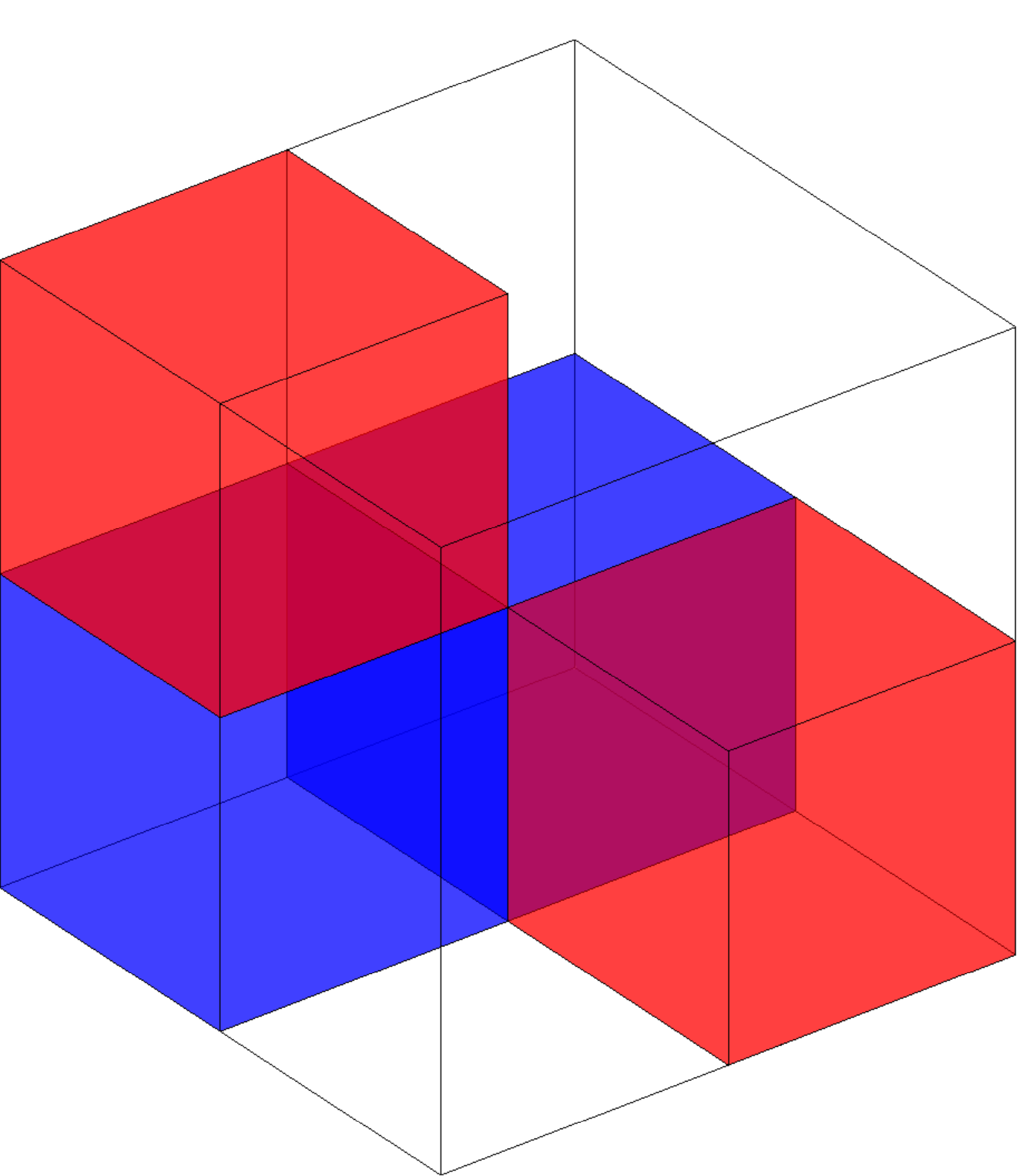}
\includegraphics[width=0.49\linewidth]{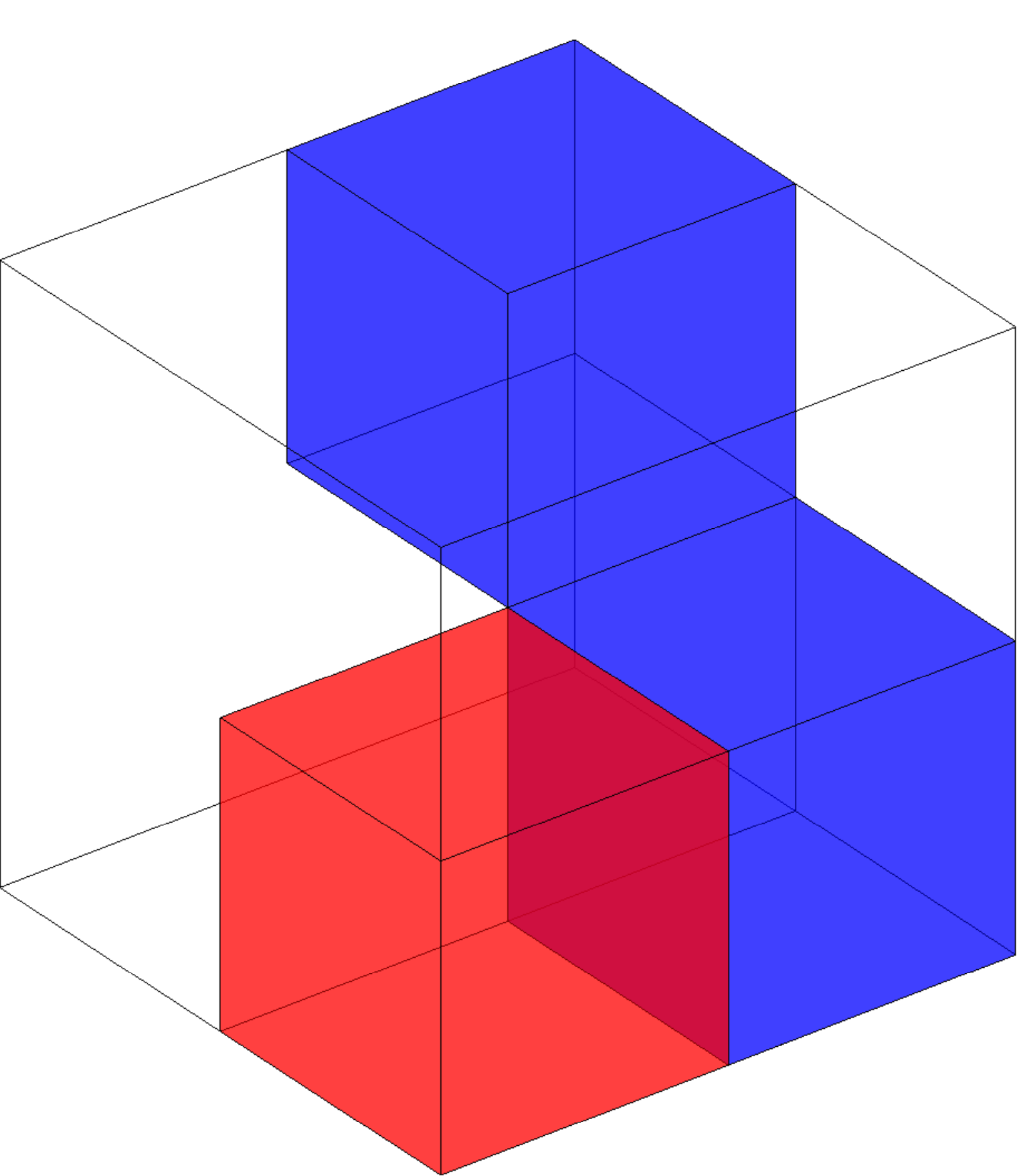}
\caption{Examples of possible configurations, where red cells are
  above the threshold, empty cells below and blue cells outside the
  survey volume. The left configuration yields $n_{0,1,2,3} = \lbrace
  \tfrac{6}{8}, \tfrac{3}{4}+1+\tfrac{2}{4},1,0 \rbrace$ and the right
  configuration yields $n_{0,1,2,3} = \lbrace \tfrac{6}{8},
  1+\tfrac{3}{4}+\tfrac{3}{4},
  1+1+\tfrac{1}{2},1\rbrace$.}
\label{figexconfs}
\end{figure}

Some examples may demonstrate this computation more clearly. Consider
the arrangement shown in figure \ref{figindexing}, assuming that the
shaded cell is above the threshold and all others are both in the
survey region and below the threshold. The contributions $n_k$ at this
node will be $\lbrace1,3,3,1\rbrace$.  Two trickier configurations
(numbers 1380 and 3647 in the sequence of $3^8$) are shown in
figure \ref{figexconfs}.

\subsection{Cell configurations for Minkowski functional contributions}

The algorithm examines the cell configuration at each node, counting
the contribution from each cell component attached to that node. We
use an unbalanced\footnote{The balanced scheme, where cells outside
  the survey are set to $-1$, is conceptually attractive but
  cumbersome to compute.} ternary labelling system, where cells within
the survey region and below the threshold are set to 0, those above
the threshold to 1 and those outside the survey region to 2. There
are, therefore, $3^8$ possible configurations for the cell values
about each node, which are enumerated by the following extension of
the Weinberg (1988) scheme:
\begin{eqnarray}
s_1 & = & 3^3(1,1,1) + 3^2(1,1,2) +  3(1,2,2) + (1,2,1) \nonumber \\
s_2 & = & 3^3(2,1,1) + 3^2(2,1,2) +  3(2,2,2) + (2,2,1)\nonumber \\
s & = & 3^4s_1 + s_2 + 1,
\end{eqnarray}
where $(i,j,k)$ is shorthand for the value of the corresponding cell
within the $2^3$ block surrounding the node. In the following we refer
to this as a function {\tt idx} mapping a $2 \times 2 \times 2$ cell
configuration to an index $s$. The cell $(1,1,1)$ is identified with
shaded cube in figure \ref{figindexing}.

\begin{algorithm}
\caption{Generate \texttt{lookup} table}
\label{alglookup}
\begin{algorithmic}
\FOR{$c = 0 \to 3^8-1$}
\STATE $n \gets \textrm{base}_3(c)$ \COMMENT{i.e., $n$ is $c$ in base 3, with 8 digits}
\STATE $A \gets \textrm{reshape}(n,[2,2,2])$ \COMMENT{pack digits into $2^3$ array}
\STATE $s \gets \texttt{idx}(A)$
\STATE $[n_0(s),n_1(s),n_2(s),n_3(s)] \gets \texttt{counts}(A)$
\ENDFOR
\end{algorithmic}
\end{algorithm}

To speed up evaluation of the total counts $N_k$ across the array, a
lookup table is used.  Algorithm \ref{alglookup} describes how this
table is generated and indexed to node configurations, using the
function {\tt counts} described in algorithm \ref{algnodecount}. Given
a three-dimensional ternary array representing the thresholded density
field, one determines the configuration at each node of the array
using the function {\tt idx} and adds the contribution $n_k(s)$ from
the table, summing these local contributions to give the total counts
$N_k$ that are the variables in equations (A1) -- (A4).

\end{document}